\documentclass[12pt]{amsart}
\usepackage[utf8]{inputenc}
\usepackage[a4paper,width=185mm,top=20mm,bottom=20mm,bindingoffset=6mm]{geometry}
\usepackage{etex}
\usepackage[all]{xy} 
\usepackage{etoolbox} 
\usepackage{lmodern}
\usepackage{array}
\usepackage{xstring}
\usepackage{longtable}
\usepackage[listings]{tcolorbox} 
\usepackage{hyperref}
\usepackage{newunicodechar}
\tcbuselibrary{breakable}
\tcbuselibrary{skins}
\usepackage{amsmath}
\usetikzlibrary{arrows.meta, positioning, shapes.geometric}
\usepackage{fancyhdr}
\usepackage{amsmath}
\usepackage{amssymb}

\usepackage{amsfonts}
\usepackage{pifont}
\usepackage[numbers]{natbib}
\usepackage{epsfig}
\usepackage{latexsym}
\usepackage{amssymb}
\usepackage{mathtools}
\usepackage{color}
\usepackage{amscd}
\usepackage{multirow}
\usepackage{graphicx}
\usepackage{lscape}
 \usepackage{tabularx}
\usepackage{float}
\usepackage{xcolor}
\usepackage{bm}
\usepackage{tikz-cd}
 \makeatletter
    \let\stdchapter\section
    \renewcommand*\section{%
    \@ifstar{\starchapter}{\@dblarg\nostarchapter}}
    \newcommand*\starchapter[1]{%
        \stdchapter*{#1}
        \thispagestyle{fancy}
        \markboth{\MakeUppercase{#1}}{}
    }
    \def\nostarchapter[#1]#2{%
        \stdchapter[{#1}]{#2}
        \thispagestyle{fancy}
    }
\makeatother

 \makeatletter
\@addtoreset{equation}{section} %\@addtoreset{equation}{Subsection}
\makeatother

\newtcolorbox{boxA}{
    fontupper = \bf,
    boxrule = 1.5pt,
    colframe = black % frame color
}

\newtheorem{theorem}{Theorem}[section]
\newtheorem*{theorem*}{Theorem}
\newtheorem{lemma}[theorem]{Lemma}
\newtheorem{proposition}[theorem]{Proposition}
\newtheorem{definition}[theorem]{Definition}
\theoremstyle{corollary}
\newtheorem{corollary}[theorem]{Corollary}
\newtheorem{remark}[theorem]{Remark}

\theoremstyle{conclusion}

\newtheorem*{Main Theorem}{Main theorem}

\usepackage{amsmath}
\usepackage{amssymb}
\usepackage{amsfonts}
\usepackage{esvect}
\usepackage{physics}
\usepackage{mathrsfs}
\usepackage{authblk}
\usepackage{geometry}
\usepackage{abstract}
\usepackage{xcolor}

\begin{document}

%\maketitle
\thispagestyle{empty}
\begin{center}
	\Large{{\bf  Equivalence of superintegrable Poisson projection chains: magnetic cotangent bundles and spin Calogero-Moser models}}
\end{center}

\vskip 0.5cm
\begin{center}
	\textsc{Kai Jiang$^{1,}$   
    Guorui Ma$^{2a,2b,\sharp}$,  %Nicolai Reshetikhin$^{3a,3b,3c,3d,\bullet}$ 
    Junze Zhang$^{3,\dagger}$} % Ian Marquette$^{3,\bullet}$,%and Yao-Zhong Zhang$^{5,\ddagger}$}
\end{center}
\begin{center}
	$^1$  School of Mathematical Sciences, Beijing University of Chemical Technology, Beijing, China
\end{center} 
%\begin{center}
%	$^2$ Department of Mathematical and Physical Sciences, La Trobe University, Bendigo, VIC 3552, Australia
%\end{center}
\begin{center}
	$2^a$ Shanghai Institute for Mathematics and Interdisciplinary Sciences (SIMIS), Shanghai, 200433, China  \\
    $2^b$  Research Institute of Intelligent Complex Systems, Fudan University, Shanghai, 200433, China 
\end{center}

%\begin{center}
%    $3^a$ Department of Mathematics, Yau Center for Mathematical Sciences, Tsinghua University, Beijing, China\\
%    $3^b$ Beijing Institute for Mathematical Sciences and Applications, Beijing, China\\
%    $3^c$ St. Petersburg State University, St. Petersburg, Russia
%    $3^d$ Department of Mathematics, University of California, Berkeley, USA
%\end{center}

\begin{center}
	$^3a$ School of Mathematics and Physics, The University of Queensland, Brisbane, QLD 4072, Australia \\
    	$^3b$ Department of Mathematical Sciences, Tsinghua University, Beijing, China
\end{center}

\begin{center}
	\footnotesize{ $\star$\textsf{kai.jiang.math@gmail.com} \hskip 0.25cm
    $\sharp$\textsf{mgr18@tsinghua.org.cn} \hskip 0.25cm
    %$^\bullet$\textsf{reshetik@math.berkeley.edu} \hskip 0.25cm
 $^\dagger$\textsf{junze.zhang@uq.net.au} 
 }
\end{center}
\vskip  0.5cm
%\hrule

\quad

\begin{abstract}
\vskip 0.15cm
\noindent
We study superintegrable Hamiltonian systems in terms of Poisson projection chains and their (spectral) equivalence. We compare two reduction constructions: (i) the cotangent bundle $T^*(G/A)$, $A=Z_G(\gamma)$, with the magnetic symplectic form defined by the orbit $G\cdot \gamma\cong G/A$, and (ii) the left-right Hamiltonian reduction of $T^*G$ that gives spin Calogero-Moser type systems. We prove that the two resulting chains are equivalent on an open set of regular points of principal orbit type: the phase spaces, the intermediate Poisson manifolds and the base spaces are identified compatibly with the projections. In particular, both constructions give the same commuting Hamiltonians.
\end{abstract}
\vskip 0.35cm

\noindent\textbf{Keywords:} Poisson geometry; superintegrable systems; Hamiltonian reduction; magnetic cotangent bundles; spin Calogero-Moser systems; coadjoint orbits.\\
\textbf{MSC 2020:} 53D17; 53D20; 37J35; 70H06.
%\hrule

\tableofcontents

\section{Introduction}
\label{sec:introd}
%Superintegrability is most naturally encoded not merely by a list of first integrals but by a nested Poisson-quotient structure. 
Integrable and superintegrable systems play an important role in the modern theory of mathematical physics \cite{MR3119484,MR3493688,MR187763,nehorovsev1968action,MR3942135,MR4071113}. A Liouville integrable Hamiltonian system on a $2n$-dimensional symplectic manifold has $n$ functionally independent first integrals in involution. A strictly superintegrable system has more than $n$ functionally independent first integrals, which cannot all be mutually commuting on a generic $2n$-dimensional symplectic leaf. Here ``superintegrable'' is used in the broad sense of noncommutative (degenerate) integrability, which includes the Liouville case \cite{MishchenkoFomenkoGeneralized1978,nehorovsev1968action}.

 Following the viewpoint of degenerately integrable systems and Poisson projection chains developed in \cite{MR3493688,MR4285685,jiang2026poisson}, we describe superintegrability on a $2n$-dimensional symplectic manifold $(M_{2n},\omega)$ by means of a superintegrable Poisson chain.
%we regard a superintegrable Hamiltonian system as data of a $2n$-dimensional symplectic manifold $(M_{2n},\omega)$ together with Poisson chain
 %\[\mathcal{M} \stackrel{\pi_1}{\longrightarrow} \mathcal{P} \stackrel{\pi_2}{\longrightarrow} \mathcal{B}.\]
 %where $\pi_1$ and $\pi_2$ are Poisson submersion, $\mathcal{P}$ is a $2n-k$-dimensional Poisson manifold, $\mathcal{B}$ is a $k$-dimensional trivial Poisson manifold. 

 \begin{definition}
\label{def:superchain1}
Let $(M_{2n},\omega)$ be a $2n$-dimensional symplectic manifold. A superintegrable system on $(M_{2n},\omega)$ is a superintegrable Poisson projection chain of integer rank $k$, with $0\leq k\leq n$, on $(M_{2n},\omega)$ that consists of Poisson manifolds $\mathcal{P}_{2n-k}$ and $\mathcal{B}_k$, and Poisson maps
\begin{equation*}
M_{2n} \xrightarrow{\ \pi_1\ } \mathcal{P}_{2n-k} \xrightarrow{\ \pi_2\ } \mathcal{B}_k, \label{eq:pmap}
\end{equation*} such that: 
\begin{enumerate}
\item $\mathcal{P}_{2n-k}$ is a $(2n-k)$-dimensional Poisson manifold, and $\pi_1: M_{2n}\to \mathcal{P}_{2n-k}$ is a surjective Poisson submersion;
\item $\mathcal{B}_{k}$ is a $k$-dimensional trivial Poisson manifold, and $\pi_2: \mathcal{P}_{2n-k} \to \mathcal{B}_{k}$ is a surjective Poisson submersion;
\item The connected components of the fibers of $\pi_2$ are symplectic leaves of $\mathcal{P}_{2n-k}$.
\end{enumerate}
The indices in the notation record the dimensions of the manifolds involved. The symplectic leaves of $\mathcal P_{2n-k}$ have dimension $2n-2k$. The case $k=n$ is the Liouville case; the term \emph{strictly superintegrable} is reserved for $k<n$.
\end{definition}

%Note that the generic fibers of $\pi_1$ have dimension $k$. If, moreover, on a dense open regular locus $\mathcal{P} \subset \mathcal{P}_{2n-k}$, a fiber of $\pi_2$ is a symplectic leaf of the Poisson manifold $(\mathcal{P}_{2n-k},\pi_P)$, then \begin{align*}
%\mathrm{rank}(\pi_P) = \dim \mathcal{P}_{2n-k} - \dim \mathcal{B}_k = (2n-k)-k = 2n-2k
%\end{align*} on $\mathcal{P}$, and the symplectic leaves have dimension $2n-2k$.

In this setting, the pullback algebra $\pi_1^*C^\infty(\mathcal{P}_{2n-k})$ records the full family of generally noncommuting first integrals, whereas the smaller algebra $(\pi_2\circ \pi_1)^*C^\infty(\mathcal{B}_k)$ records the commuting Hamiltonians. Since $\pi_1$ and $\pi_2$ are surjective submersions, we may regard $C^\infty(\mathcal P_{2n-k})$ and $C^\infty(\mathcal B_k)$ as subalgebras of $C^\infty(M_{2n})$. For a quantum counterpart, in which centralizer algebras play the role of $\pi_1^*C^\infty(\mathcal P_{2n-k})$, see \cite{MR4700605}.

%We also have the algebraic description of a superintegrable system. In other words, Poisson commutative subalgebra $C_{I}(\mathcal{M}) \subset C(\mathcal{M})$ defines a superintegrable system if its Poisson centralizer $C_{J}(\mathcal{M})$ has rank $2n-k$.

% \begin{definition}   A superintegrable system on a $2n$-dimensional symplectic manifold $(M,\omega)$ consists of a Poisson subalgebra $C_{J}(M) \subset C^{\infty}(M)$ of rank $2n-k$ that has a Poisson center $C_{I}(M)$ of rank $k$.  \end{definition}

\medskip
\noindent\textbf{Standing convention on the Lie group.}
Unless explicitly stated otherwise, $G$ denotes a compact, connected, semisimple Lie group with Lie algebra $\mathfrak{g}$. For the magnetic models, fix $\gamma\in\mathfrak{g}$ and a real parameter $\varepsilon\in\mathbb R$. Write $A:=Z_G(\gamma)$ and $\mathfrak a:=\mathrm{Lie}(A)$. The space $\widetilde{\mathfrak a}:=\mathrm{Ann}(\mathfrak a)\subset\mathfrak{g}^*$ consists of covectors vanishing on $\mathfrak a$. Let $B$ be the Killing form, put $\lambda:=B(\gamma,\cdot)\in\mathfrak{g}^*$, and let $\pi_{\mathfrak a}:\mathfrak{g}^*\to\mathfrak a^*$ denote restriction.

\medskip
In this paper, we focus on the following two superintegrable systems. For more examples of superintegrable systems, see the survey paper \cite{MR3493688}.

\medskip
The first one is the magnetic flow on a reductive homogeneous space \cite{Bolsinov2008,jiang2026poisson}, where twisted (magnetic) cotangent bundles are studied and one often uses the left trivialization $T^*G\cong G\times\mathfrak{g}^*$ to write the relevant phase spaces in a convenient $(G\times\mathfrak{g}^*)$-type model (or, for $T^*(G/A)$, an associated-bundle model $G\times_A\widetilde{\mathfrak{a}}$). In this framework, the Hamiltonians and noncommutative first integrals are organized into explicit polynomial Poisson algebras, and the integrable family is obtained by taking spectral invariants (e.g., coefficients of characteristic polynomials) of the $\mathfrak{g}^*$-valued moment map.

\medskip
The second class of superintegrable systems is the double spin Calogero-Moser model \cite{MR3493688,MR4285685,jiang2026poisson}.
%, in which a Lie group $G$ and its Lie subgroups $K_1,K_2\subset G$, Calogero-Moser type Hamiltonian systems on the double quotient, or symmetric-space phase space $K_1\backslash T^*G/K_2$, were constructed. These models
For a Lie group $G$ and Lie subgroups $K_1,K_2\subset G$, Calogero-Moser type Hamiltonian systems on the double-quotient phase space $K_1\backslash T^*G/ K_2$ were constructed in \cite{MR4285685}. These models arise from the cotangent bundle $T^*G$ with its canonical symplectic structure, together with the natural left-right action of $K_1\times K_2$ and a reduction procedure. Under standard regularity assumptions on the $K_1\times K_2$-orbit structure, the reduced spaces are Poisson or symplectic manifolds, and their first integrals form a polynomial Poisson algebra containing a Poisson-commutative subalgebra of Hamiltonians.

\medskip

Our aim is to compare the two constructions at the level of Poisson projection chains. Following \cite{MR3493688}, we use two notions. A spectral equivalence consists of Poisson maps between the corresponding spaces of the two chains, commuting with the projections, such that the map between the bases is a diffeomorphism. An equivalence is a spectral equivalence in which all three maps are diffeomorphisms. Only the latter is an equivalence relation in general.

\begin{definition} 
\label{def:equidefi}
Let $(M_{2n},\omega)$ and $(M_{2n'}',\omega')$ be symplectic manifolds % of different dimension  
such that 
\begin{align}
M_{2n}  \xrightarrow{\ \pi_1\ } \mathcal{P}_{2n-k}\xrightarrow{\ \pi_2\ } \mathcal{B}_k \  \text{ and } \  M_{2n'}' \xrightarrow{\ \pi_1'\ } \mathcal{P}_{2n'-k'}' \xrightarrow{\ \pi_2'\ } \mathcal{B}_{k'}'  \label{eq:twosystem}
\end{align} are superintegrable.
The systems in \eqref{eq:twosystem} are called \emph{spectrally equivalent} if there exist Poisson maps \begin{align*}
\phi:M_{2n} \rightarrow  M_{2n'}',\qquad \phi_1:\mathcal{P}_{2n-k}\rightarrow  \mathcal{P}_{2n'-k'}',\qquad \phi_2:\mathcal{B}_k \rightarrow  \mathcal{B}_{k'}'
\end{align*}
such that $\phi_2$ is a diffeomorphism and the following diagram commutes:
\[\begin{tikzcd}
	M_{2n} & {\mathcal{P}_{2n-k}} & {\mathcal{B}_k} \\
	{M_{2n'}'} & {\mathcal{P}_{2n'-k'}'} & {\mathcal{B}_{k'}'}
	\arrow["\pi_1", from=1-1, to=1-2]
	\arrow["{\phi}", from=1-1, to=2-1]
	\arrow["\pi_2", from=1-2, to=1-3]
	\arrow["{\phi_1}", from=1-2, to=2-2]
	\arrow["{\phi_2}", from=1-3, to=2-3]
	\arrow["{\pi_1'}", from=2-1, to=2-2]
	\arrow["{\pi_2'}", from=2-2, to=2-3]
    \arrow[from=1-1, to=2-2, phantom, "\circlearrowright" {anchor=center, scale=1.5, rotate=90}]
    \arrow[from=1-2, to=2-3, phantom, "\circlearrowright" {anchor=center, scale=1.5, rotate=90}]
\end{tikzcd}\]
Equivalently, the commutativity conditions are
\begin{align*}
\pi_1'\circ \phi=\phi_1\circ \pi_1,\qquad \pi_2'\circ \phi_1=\phi_2\circ \pi_2.
\end{align*}
They are called \emph{equivalent} if, moreover, $\phi$ and $\phi_1$ are diffeomorphisms.
\end{definition} % In particular, our notion of \emph{spectral equivalence} isolates the part of the dynamics that is encoded by the spectral invariants and provides a practical criterion for deciding when two constructions yield the same commutative Hamiltonians and the same noncommutative algebra of first integrals up to Poisson maps.
%This allows one to compare superintegrable systems produced by different reduction procedures in a uniform language and clarifies which features are intrinsic to the system and which depend on the presentation.
%For a Lie group $G$ and two subgroups $K_1, K_2 \subset G$, we consider a $(K_1 \times K_2)$-action on $T^*G$. There is a moment map $\mu_{12}:T^*G \longrightarrow \mathfrak{k}_1^* \times \mathfrak{k}_2^*$ for the $K_1 \times K_2$-action where $\mathfrak{k}_1^*,\mathfrak{k}_2^*$ is the corresponding dual Lie algebra of $K_1,K_2$. For any two coadjoint orbits $\mathcal{O}_1 \subset \mathfrak{k}_1^*$ and $\mathcal{O}_2 \subset \mathfrak{k}_2^*$, A phase space $S(\mathcal{O}_1,\mathcal{O}_2)$ can be constructed by using the pull-back of $\mu_{12}$ and the reduction of the $K_1 \times K_2$-action. A superintegrable Poisson projection chain on this phase space $S(\mathcal{O}_1, \mathcal{O}_2)$ can be constructed, see Proposition \ref{prop:superintchainso1o2}. Our focus is on comparing such superintegrable constructions at the level of their Poisson projection chains. Building on the polynomial Poisson algebras of first integrals arising in reductions such as \cite{MR4285685} (and their twisted counterparts), we introduce and study equivalence relations, in particular, spectral equivalence, that formalize when two superintegrable systems have the same base commutative Hamiltonians and the same noncommutative algebraic structure up to Poisson maps. In other words, 
Equivalent chains describe the same superintegrable system up to a Poisson change of variables. Our main result, Theorem \ref{thm:specequiv}, states that for $A=Z_G(\gamma)$ and $(K_1,K_2)=(\{e\},A)$, reduction of $T^*G$ at $(0,\varepsilon\pi_{\mathfrak a}(\lambda))$ gives a chain equivalent to the magnetic chain on $T^*(G/A)$ on an open set of regular points of principal orbit type. When $\gamma$ is regular, so that $A=T$, a further reduction by the left $T$-action gives a superintegrable $(T,T)$-chain (Subsection \ref{subsec:coisotropy}). Related cotangent bundle reductions for polar actions and spin Calogero-Moser models are studied in \cite{Hochgerner2008SingularCotangent}; see also \cite{Feher2023Master,Feher2024Degenerate} for Poisson reductions of free motion systems on compact Lie groups.

The notion of regularity plays an important role in this paper. We now distinguish these notations by means of the following:

\begin{enumerate}
    \item Let $M$ and $N$ be smooth manifolds and $F: M \rightarrow N$ be a smooth map. The \textit{maximal-rank locus} of $F$ is
\begin{align*}
M_\mathrm{reg}(F):=\bigl\{m\in M:\mathrm{rank} \, dF_m = \max_{z\in M}\mathrm{rank} \, dF_z\bigr\}.
\end{align*}
This should not be confused with the notion of a \emph{regular value}: a point $y\in N$ is a regular value if $dF_m$ is surjective for every $m\in F^{-1}(y)$ (equivalently, $\mathrm{rank}\,dF_m=\dim N$ on the fiber). If the maximal rank of $F$ equals $\dim N$, then the maximal-rank locus is precisely the set of regular points; moreover, $y$ is a regular value if and only if every point of the fiber $F^{-1}(y)$ is a regular point.

If $F$ has a constant rank on an open subset $U\subset M$ (in particular, if $F|_U$ is a submersion onto its image), then, by the constant rank theorem, the fibers of $F|_U$ form a smooth family.

    \item For a Poisson manifold $(P,\pi_P)$, the \textit{Poisson regular locus} is
\begin{align*}
P_\mathrm{reg}:=\bigl\{p\in P:\mathrm{rank} \, \pi_P(p)=\max_{q\in P}\mathrm{rank} \, \pi_P(q)\bigr\}.
\end{align*}
It is the union of the symplectic leaves of maximal dimension.

 \item For the Lie algebra $\mathfrak{g}$, the \textit{adjoint regular locus} is
\begin{align}
\mathfrak{g}_\mathrm{reg}:=\bigl\{X \in\mathfrak{g}:\dim Z_G(X) = \mathrm{rank} \, G\bigr\}.
\label{eq:regular}
\end{align}
%Equivalently, $X\in\mathfrak{g}_\mathrm{reg}$ if and only if its adjoint orbit has maximal dimension. 
Under a fixed identification $\mathfrak{g}\cong\mathfrak{g}^*$, this is equivalent to requiring that the Lie-Poisson tensor have maximal rank. Similarly, we can define the \textit{coadjoint regular locus} $\mathfrak{g}_\mathrm{reg}^*$ on the dual algebra.
\end{enumerate}

Note that the adjoint regularity is a statement about the \emph{$G$-orbit} of $X$ and does \emph{not} imply freeness (or even minimal stabilizer) for an additional subgroup action (e.g., the $A$-action on the affine slice $\mathfrak{m}-\varepsilon \gamma$). Hence, we use two subscripts for two different purposes. For subsets of $\mathfrak{g}$ and $\mathfrak{g}^*$, the subscript ``$\mathrm{reg}$'' denotes adjoint and coadjoint regularity, respectively. The notations $M_{\mathrm{reg}}(F)$ and $P_{\mathrm{reg}}$ retain the differential-rank and Poisson-rank meanings defined above.

The spaces in our projection chains are obtained by \emph{imposing constraints} (via moment maps) and then \emph{quotienting by additional subgroup actions}. (Co)adjoint regularity does not, by itself, ensure constant rank for the relevant maps, nor constant orbit type for the subgroup actions used in the quotients. These are the hypotheses that guarantee that the reduced spaces and the fiber products used to define intermediate Poisson manifolds are smooth.
Throughout this paper, the subscript $\mathrm{pr}$ means the principal orbit-type locus of the specified group action. For a proper action of a compact group $K$ on a connected manifold $N$, this is
\begin{align*}
N_{\mathrm{pr}}:=\{z\in N:K_z\text{ is conjugate in }K\text{ to }H\},
\end{align*}
where $K_z:=\{k\in K:k\cdot z=z\}$ and $H$ is a principal stabilizer. Equivalently, $N_{\mathrm{pr}}$ is the unique orbit-type subset that is open and dense in $N$. Coadjoint regularity, constant-rank restrictions, and effective freeness are stated separately.

% In this work, $G$ is a real, connected, compact Lie group acting linearly on a real affine space $X$ (in particular, on subspaces of $\mathfrak{g}$). We write the affine GIT quotient as \begin{align*}
%  X//G:=\mathrm{Spec}\bigl(\mathbb{R}[X]^G\bigr),   
% \end{align*} so the quotient map $\chi:X\to X//G$ is the map given by $G$-invariant polynomial functions. The notation $\mathrm{Spec}$ is kept only to record the invariant polynomial coordinates.

% In particular, for the coadjoint representation we obtain the (co)adjoint quotient map $\chi:\mathfrak{g}^*\to \mathfrak{g}^*//G$. For compact $G$, the invariant polynomials separate coadjoint orbits, and the role of restricting to $\mathfrak{g}_\mathrm{reg}^*$ in this paper is that $\chi$ has maximal rank there and its fibers have constant orbit type; hence, for $\xi,\eta\in\mathfrak{g}_\mathrm{reg}^*$, we have $\chi(\xi)=\chi(\eta)$ if and only if $G\cdot\xi = G\cdot\eta$.

\section{Equivalence between superintegrable chains}
\label{sec:tgdoubleequivalent}

 The purpose of this section is to establish the equivalence between the following two superintegrable Poisson projection chains in the sense of Definition \ref{def:equidefi}.
The first one is on the twisted cotangent bundle chain in Theorem \ref{thm:superin}, and the second one is the specialization $K_1 = \{e\}, K_2 = A$,  $(\mathcal{O}_1,\mathcal{O}_2) = (\{0\},\{\varepsilon\pi_{\mathfrak{a}}(\lambda)\})$ of the general left-right Hamiltonian reduction chain on $T^*G$ recalled from \cite{MR4285685}. We begin by recalling some basic terminology applied on constructing the twisted cotangent bundle chain. Throughout this section, $G$ is compact, connected, and semisimple, and  $B$ denotes the Killing form, %Fix the $\mathrm{Ad}(G)$-invariant nondegenerate bilinear form $B$ 
and write \begin{align*}
B^\flat:\mathfrak{g} \longrightarrow \mathfrak{g}^*,\quad B^\flat(X):=B(X,\cdot), \quad B^\sharp :=(B^\flat)^{-1}.
\end{align*}
%For an affine subspace $E\subset\mathfrak{g}$, we write  $E^\flat:=B^\flat(E) \subset\mathfrak{g}^*$.
The left-right Hamiltonian reduction chain is naturally written in intrinsic dual notation: moment maps take values in dual Lie algebras, the relevant orbits are coadjoint orbits, and equivariance is expressed using $\mathrm{Ad}^*$. %We use the same dual notation for the magnetic cotangent-bundle chain in Theorem \ref{thm:superin} .

We use the convention that for a (finite-dimensional) vector space $W$, $S(W)$ denotes its symmetric algebra, identified with the algebra of polynomial functions on $W^*$. For an affine subspace $E\subset\mathfrak{g}^*$, we write $\mathbb{R}[E]$ for its algebra of polynomial functions.

We use the coadjoint quotient (affine GIT quotient) map \begin{align}
\chi :\mathfrak{g}^*\longrightarrow\mathfrak{g}^*// G:=\mathrm{Spec}\bigl(S(\mathfrak{g})^G\bigr), \label{eq:affinegit}
\end{align}
%Via $B^\flat$, these quotient spaces are canonically identified, and $ \chi_{\mathfrak{g}^*}(B^\flat X)=\chi_{\mathfrak{g}}(X)$ after this identification.
Here $//$ is the affine GIT quotient, used only to name invariant-polynomial coordinates. Later, $P/G$ denotes an orbit space, and $//_\xi$ symplectic reduction at level $\xi$. All manifolds in our chains are open subsets of orbit-type strata of such orbit spaces. On $\mathfrak{g}^*_{\mathrm{reg}}$, $\chi$ is a submersion onto the manifold of regular coadjoint orbits.

 For a fixed element $\gamma\in\mathfrak{g}$, let
\begin{align*}
    A:=Z_G(\gamma),\qquad \mathfrak{a}:=\mathrm{Lie}(A),\qquad \mathfrak{g}=\mathfrak{a}\oplus \mathfrak{m}
\end{align*}
be the $B$-orthogonal vector space decomposition, and set $M:=G/A$. Set $\lambda:=B^\flat(\gamma) \in \mathfrak{g}^*$. For any Lie subalgebra $\mathfrak{k}\subset \mathfrak{g}$, let $\pi_{\mathfrak{k}}:\mathfrak{g}^*\to\mathfrak{k}^*$ be the restriction and put \begin{align*}
\widetilde{\mathfrak{a}}:=\ker\pi_{\mathfrak{a}}=B^\flat(\mathfrak{m}),\qquad \widetilde{\mathfrak{a}}-\varepsilon\lambda=B^\flat(\mathfrak{m}-\varepsilon \gamma).
\end{align*}
Here $\widetilde{\mathfrak{a}}\subset\mathfrak{g}^*$ is the annihilator of $\mathfrak{a}\subset\mathfrak{g}$, i.e., the covectors vanishing on $\mathfrak{a}$. Via the decomposition $\mathfrak{g}=\mathfrak{a}\oplus\mathfrak{m}$, restriction to $\mathfrak{m}$ identifies $\widetilde{\mathfrak{a}}\cong\mathfrak{m}^*\cong(\mathfrak{g}/\mathfrak{a})^*$.
The coadjoint stabilizer $G_\lambda$ equals $A$. %Thus this is the same magnetic model, with both chains expressed in covectors.  
Fix a maximal torus $T\subset A$ and set $\mathfrak{t}:=\mathrm{Lie}(T)$.
Note that $Z(G)\subseteq A$ and the coadjoint action of $A$ on $\widetilde{\mathfrak{a}}$ (hence on $\widetilde{\mathfrak{a}}-\varepsilon\lambda$) have a kernel containing $Z(G)$. Quotienting by this kernel makes the coadjoint action effective but does not, in general, make it free. The freeness is required only on the loci where it is used. %The principal right action on $G$ is still by $A$.

We use two conventions for the action of $A$ on $G$. As a principal bundle, $G\to G/A =: M$ carries the right action $g\cdot a=ga$. The connections and the associated bundle $G\times_A\widetilde{\mathfrak a}$ refer to this action. Hamiltonian reduction requires a left action of $K_1\times K_2$, so the second factor acts by $a\cdot g=ga^{-1}$. Its infinitesimal generator is $-(dL_g)_e\zeta$, the negative of that of the principal action.
We use the right trivialization $T^*G\simeq G\times\mathfrak{g}^*$ defined by
\begin{equation}
\alpha_g(v)=\langle\xi,(dR_{g^{-1}})_g v\rangle,\qquad v\in T_gG. \label{eq:right-trivialization}
\end{equation}
Here $\langle\xi,X\rangle$ is dual evaluation, while $B(X,Y)$ denotes the fixed bilinear form. Thus, $\langle B^\flat X,Y\rangle=B(X,Y)$. Using $\widetilde{\mathfrak{a}}\cong(\mathfrak{g}/\mathfrak{a})^*$, we identify\begin{align*}
T^*(G/A)\cong G\times_A  \widetilde{\mathfrak{a}}
 \end{align*} and we write a point of $T^*M$ as $(g,X):=[g,X]_A$ with $g\in G$ and  $X\in\widetilde{\mathfrak{a}}$, where \begin{align*}
[g,X]_A = [ga,\mathrm{Ad}_{a^{-1}}^* X]_A \qquad \text{for all }a\in A.
\end{align*}
Let $\iota_\lambda : M \to G\cdot\lambda$, $\iota_\lambda(gA)=\mathrm{Ad}_g^*\lambda$ identify $M$ with the coadjoint orbit of $\lambda$, and let $\omega_{\mathrm{KKS},\lambda}$ be the Kostant-Kirillov-Souriau (KKS) form on this orbit, following the sign convention of Appendix \ref{app:twistedform}. We define \begin{align*}
\sigma_{\gamma}:=-\iota_\lambda^*\omega_{\mathrm{KKS},\lambda}\in\Omega^2(M).
\end{align*}
With this sign, the left $G$-action on $\bigl(T^*M,\omega_{\mathrm{can}}+\varepsilon\,\pi^*\sigma_{\gamma}\bigr)$ has moment map $J_G([g,X]_A)=\mathrm{Ad}_g^*(X-\varepsilon\lambda)$. Here $X$ is the coordinate: the covector $[g,X]_A$ pulls back to $T_g^*G$ as $v\mapsto\langle X,(dL_{g^{-1}})_g v\rangle$.

We now choose smooth domains to construct the projection chain with the phase space $T^*(G/A)$. For the regular-locus comparison, assume explicitly that
\begin{align*}
(\widetilde{\mathfrak{a}}-\varepsilon\lambda)\cap\mathfrak{g}_\mathrm{reg}^*\neq\varnothing. \tag{R}\label{eq:nonempreg}
\end{align*}
%This condition is a genuine restriction. %For $G=\mathrm{SU}(4)$ and $\gamma=i\mathrm{diag}(1,1,1,-3)$, corresponding to $G/A=\mathbb{CP}^3$, each element of $\mathfrak m-\varepsilon \gamma$ has the form $i\left(\begin{smallmatrix}-\varepsilon I_3&v\\v^*&3\varepsilon\end{smallmatrix}\right)$, with $v\in\mathbb C^3$. The subspace $v^\perp\oplus\{0\}$ has dimension at least two and eigenvalue $-i\varepsilon$, so there are no regular points for any $\varepsilon$. In contrast, for $\mathbb{CP}^2$ the eigenvalues divided by $i$ are $-\varepsilon$ and $\varepsilon/2\pm\sqrt{9\varepsilon^2/4+\|v\|^2}$, which are distinct for $v\ne0$; for $\mathrm{Gr}(2,4)$ they can be chosen as $\pm\sqrt{\varepsilon^2+s_1^2}$ and $\pm\sqrt{\varepsilon^2+s_2^2}$, with $s_1>s_2>0$. Here the defining elements are $i\mathrm{diag}(1,1,-2)$ and $i\mathrm{diag}(1,1,-1,-1)$, respectively, and $s_1,s_2$ are the singular values of the off-diagonal block. If $\gamma$ is regular and $\varepsilon\ne0$, the point $-\varepsilon\lambda$ already verifies the condition.
Consider the principal orbit-type locus for the coadjoint $A$-action on the affine subspace $\widetilde{\mathfrak{a}}-\varepsilon\lambda$, i.e., $(\widetilde{\mathfrak{a}}-\varepsilon\lambda)_{\mathrm{pr}}\subseteq \widetilde{\mathfrak{a}}-\varepsilon\lambda$. We then work on the $A$-invariant open subset
\begin{align*}
(\widetilde{\mathfrak{a}}-\varepsilon\lambda)_{\mathrm{pr}}\cap\mathfrak{g}_\mathrm{reg}^*
\subseteq (\widetilde{\mathfrak{a}}-\varepsilon\lambda)\cap\mathfrak{g}_\mathrm{reg}^*,
\end{align*}
which is nonempty since $(\widetilde{\mathfrak{a}}-\varepsilon\lambda)_{\mathrm{pr}}$ is open dense in $\widetilde{\mathfrak{a}}-\varepsilon\lambda$. For $y \in (\widetilde{\mathfrak{a}}-\varepsilon\lambda)_{\mathrm{pr}}\cap\mathfrak{g}_\mathrm{reg}^*$, write $h=\dim A_y$ as the (constant) stabilizer dimension, where $A_y=\{a\in A:\mathrm{Ad}_a^* y=y\}$, and set $k=\mathrm{rank}G-h$. Write $\mathfrak{g}_y:=\{u\in\mathfrak{g}:\mathrm{ad}_u^*y=0\}$. On $(\widetilde{\mathfrak{a}}-\varepsilon\lambda)_{\mathrm{pr}}\cap\mathfrak{g}_\mathrm{reg}^*$,
\begin{align*}
\mathrm{rank}d(\chi\vert_{  \widetilde{\mathfrak{a}} -\varepsilon \lambda })_y =\mathrm{rank}G-\dim(\mathfrak{a}\cap \mathfrak{g}_y )=k.
\end{align*}
Indeed, the differentials of the coadjoint invariants span $\mathfrak{g}_y\subset(\mathfrak{g}^*)^*$ at a regular point, and the restriction to $\widetilde{\mathfrak{a}}$ has kernel $\mathfrak{a}\cap\mathfrak{g}_y$. To obtain an embedded image under $\chi$, take a nonempty $A$-invariant open subset
\begin{align*}
V\subset(\widetilde{\mathfrak{a}}-\varepsilon\lambda)_{\mathrm{pr}}
\cap\mathfrak{g}^*_{\mathrm{reg}}
\end{align*}
such that $\chi(V)$ is an embedded $k$-dimensional submanifold of $\chi(\mathfrak{g}^*_{\mathrm{reg}})$. %Such a subset is obtained by saturating a sufficiently small constant-rank neighborhood under the compact group $A$. Then $\chi|_V$ is a submersion onto its image. %The notation $(\widetilde{\mathfrak{a}}-\varepsilon\lambda)_{\mathrm{pr}}$ continues to denote the entire principal orbit-type locus. 
Fix this choice of $V$ and define
\begin{align}\label{eq:comparison-domains}
U_0&:=\chi(V),\\
U&:=\{[g,X]_A:X-\varepsilon \lambda\in V\},\\
U_1&:=\{(x,[y]_A):x\in\chi^{-1}(U_0)\cap\mathfrak{g}_\mathrm{reg}^*,\
 y\in V,\ \chi(x)=\chi(y)\}.
\end{align}
The orbit space $V/A$ is smooth due to the constant orbit-type condition. Writing $n:=\dim G$ and $h:=\dim A_y$ on $V$, we have \begin{align*}
\dim(V/A)=\dim V-(\dim A-h)=(d-\dim A)-(\dim A-h) = n-2\dim A+h.
\end{align*} Moreover, by construction $\dim U_0=\dim\chi(V)=k$, the fiber product description gives $\dim U_1=\dim U-k$.
Both maps to $U_0$ are submersions, so $U_1$ is their smooth fiber product. Define $\Pi_1: U\to U_1$ and $\Pi_2 :  U_1\to U_0$ by \begin{align*}
\Pi_1([g,X]_A) =(\mathrm{Ad}_g^* (X-\varepsilon \lambda ),[X-\varepsilon \lambda ]_A),\qquad \Pi_2(x,[y]_A) =\chi(y).
\end{align*} These choices remain fixed throughout Theorems \ref{thm:superin} and \ref{thm:specequiv}. %Since $\chi$ separates compact coadjoint orbits, we immediately have that $\Pi_1$ is surjective. Its differential is surjective: before quotienting, any compatible tangent pair $(\dot x,\dot y)$ satisfies   $\dot x-\mathrm{Ad}_g^*\dot y\in\ker d\chi_x=T_x(G\cdot x)$, and hence comes from varying $g$ and $y$. Thus, the required nonempty smooth loci exist under the stated non-emptiness hypothesis.% These are local regular-locus statements.
The following theorem is proved in \cite{jiang2026poisson}.
\begin{theorem}
  \label{thm:superin} 
Let $G$ be a compact, connected, semisimple Lie group, and let $A=Z_G(\gamma)$. Let $G/A$ be the reductive homogeneous space. Fix $\varepsilon\in\mathbb{R}$. Equip $T^*M$ with the magnetic symplectic form $\omega_\varepsilon:=\omega_{M,\mathrm{can}}+\varepsilon\,\pi_M^*\sigma_{\gamma}$ determined by $\gamma$, where $\pi_M:T^*M\to M$ is the bundle projection. Assume that the condition \eqref{eq:nonempreg} holds.
Then the restricted chain %\footnote{We write $\Pi_1,\Pi_2$ here to avoid confusion with the natural projections $\pi_i:\mathfrak{g}^*\to\mathfrak{k}_i^*$ used in the definition of $S(\mathcal{O}_1,\mathcal{O}_2)$.} 
\begin{align*}
U\xrightarrow{\,\Pi_1\,}U_1\xrightarrow{\,\Pi_2\,}U_0
\end{align*}
is a superintegrable Poisson projection chain.
\end{theorem}

\subsection{Zero-level biquotient reduction and the choice of coadjoint orbits}
\label{subsec:zerolevelbiquotient}

Let $K_1,K_2\subset G$ be closed Lie subgroups with Lie algebras $\mathfrak{k}_1,\mathfrak{k}_2$.
Throughout this subsection, we assume that the left-right action of $K_1\times K_2$ on $G$,
\begin{align*}
(k_1,k_2)\cdot g := k_1 g k_2^{-1},
\end{align*}
is free and proper, so that the biquotient $Q:=K_1\backslash G/K_2$ is a smooth manifold. (Notice that, under this assumption, the cotangent lift of the action on $T^*G$ is also free and proper.)

Using the right trivialization \eqref{eq:right-trivialization} and the convention $\omega_{\mathrm{can}}=-d\vartheta_{G}$, where $\vartheta_{G}$ is the tautological one-form on $T^*G$, the left and right moment maps are \begin{align*}
\mu_L(g,\xi)=\xi,\qquad \mu_R(g,\xi)=-\mathrm{Ad}_{g^{-1}}^*\xi .
\end{align*} Thus, for the $K_1\times K_2$-action,
\begin{align}
\mu_{12}(g,\xi) = \left( \pi_1(\xi), -\pi_2(\mathrm{Ad}_{g^{-1}}^*\xi) \right) \in \mathfrak{k}_1^*\oplus\mathfrak{k}_2^*.
\label{eq:coorlrmp}
\end{align}
where $\pi_i:\mathfrak{g}^*\to\mathfrak{k}_i^*$ are the natural projections.

Let $\mathcal{O}_i\subset\mathfrak{k}_i^*$ be coadjoint orbits. We write the full reduced space as
\begin{align*}
S(\mathcal{O}_1,\mathcal{O}_2) := K_1\backslash \mu_{12}^{-1}(\mathcal{O}_1\times\mathcal{O}_2)/K_2.
\end{align*}
For the comparison arguments in this section, we further restrict to the coadjoint-regular piece
\begin{align*}
S^{\mathrm{reg}}(\mathcal{O}_1,\mathcal{O}_2) := K_1\backslash \left\{ (g,\xi)\in G\times\mathfrak{g} _\mathrm{reg}^* : \pi_1(\xi)\in\mathcal{O}_1,\, -\pi_2(\mathrm{Ad}_{g^{-1}}^*\xi)\in\mathcal{O}_2 \right\}/K_2,
\end{align*}
whose connected components are symplectic leaves of the Poisson manifold $K_1\backslash (G\times\mathfrak{g}^*_{\mathrm{reg}})/K_2$.

Write $\mathfrak{g}^*\times\mathfrak{g}^*_-$ for $\mathfrak{g}^*\times\mathfrak{g}^*$ with the Lie-Poisson structure on the first factor and the opposite structure on the second. The map
\begin{align*}
\widetilde p_1:=(\mu_L,-\mu_R) :  T^*G\longrightarrow\mathfrak{g}^*\times\mathfrak{g}^*_-,\qquad \widetilde p_1(g,\xi)=(\xi,\mathrm{Ad}_{g^{-1}}^*\xi),
\end{align*}
is Poisson and $K_1\times K_2$-equivariant. With $\chi$ as in \eqref{eq:affinegit}, define
\begin{align}
P(\mathcal{O}_1,\mathcal{O}_2) := \left\{ (x,y)\in\mathfrak{g}  _\mathrm{reg}^*\times\mathfrak{g}_\mathrm{reg}^* : \pi_1(x)\in\mathcal{O}_1,\, -\pi_2(y)\in\mathcal{O}_2,\, \chi(x)= \chi(y) \right\}\big/(K_1\times K_2),
\label{eq:intermide}
\end{align} with the Poisson structure induced from $\mathfrak{g}^*\times\mathfrak{g}^*_-$, and \begin{align}
B(\mathcal{O}_1,\mathcal{O}_2) := \chi\left( \left\{ y\in\mathfrak{g} _\mathrm{reg}^* : -\pi_2(y)\in\mathcal{O}_2,\  \exists x \in\mathfrak{g}_\mathrm{reg}^* \text{ with }\pi_1(x)\in\mathcal{O}_1,\ \chi(x)=\chi(y) \right\} \right), \label{eq:base}
\end{align}
endowed with the trivial Poisson structure. %These spaces are in general only stratified; 
In general, $P(\mathcal{O}_1,\mathcal{O}_2)$ and $B(\mathcal{O}_1,\mathcal{O}_2)$ are not manifolds, and the next proposition is understood on open subsets on which all spaces are smooth. Precisely, let $W\subset S^{\mathrm{reg}}(\mathcal{O}_1,\mathcal{O}_2)$ be open, put $P^\circ:=p_1(W)$ and $B^\circ:=p_2(P^\circ)$, and assume that $K_1\times K_2$ acts properly with a single orbit type on the preimage of $P^\circ$, that $B^\circ$ is an embedded submanifold of $\chi(\mathfrak{g}^*_{\mathrm{reg}})$, that $p_1|_W$ and $p_2|_{P^\circ}$ are submersions, and that $\dim W=\dim P^\circ+\dim B^\circ$. The subsets used in Theorems \ref{thm:specequiv} and \ref{thm:superinchaini} are of this kind.

\begin{proposition}\label{prop:superintchainso1o2}\cite[Section 1.3]{MR4285685}
The maps
\begin{align*}
p_1: & S^{\mathrm{reg}}(\mathcal{O}_1,\mathcal{O}_2)\to P(\mathcal{O}_1,\mathcal{O}_2),\quad [g,\xi]\mapsto[(\xi,\mathrm{Ad}_{g^{-1}}^*\xi)],\\
p_2: & P(\mathcal{O}_1,\mathcal{O}_2)\to B(\mathcal{O}_1,\mathcal{O}_2),\quad [(x,y)]\mapsto\chi(x)
\end{align*}
are well defined, and
\begin{equation}\label{eq:chain_so1o2}
S^{\mathrm{reg}}(\mathcal{O}_1,\mathcal{O}_2)\xrightarrow{\ p_1\ }P(\mathcal{O}_1,\mathcal{O}_2)\xrightarrow{\ p_2\ }B(\mathcal{O}_1,\mathcal{O}_2)
\end{equation}
is a superintegrable Poisson projection chain.
\end{proposition}

\begin{proof}
We work on $W$, $P^\circ$, $B^\circ$ as above, and write $p_1,p_2$ for the restricted maps.
The right-trivialized cotangent lift of $(k_1,k_2)\cdot g=k_1gk_2^{-1}$ is \begin{align*}
(g,\xi)\longmapsto(k_1gk_2^{-1},\mathrm{Ad}_{k_1}^*\xi),
\end{align*}
and \begin{align*}
\widetilde p_1(k_1gk_2^{-1},\mathrm{Ad}_{k_1}^*\xi) =\bigl(\mathrm{Ad}_{k_1}^*\xi,\mathrm{Ad}_{k_2}^*\mathrm{Ad}_{g^{-1}}^*\xi\bigr).
\end{align*}
. Thus, $\widetilde p_1$ is equivariant and descends to $p_1$. Since $\widetilde p_1$ is Poisson, its descended map is compatible with Poisson reduction. In particular, it is a smooth Poisson map on $P^\circ$.

Every constrained pair $(x,y)$ with $\chi(x)=\chi(y)$ satisfies $x=\mathrm{Ad}_g^*y$ for some $g\in G$, because compact coadjoint orbits are separated by $\chi$. The point $(g,x)$ belongs to the required moment-map constraint set and maps to $[(x,y)]$. Thus, $p_1$ is surjective. Moreover, $p_2$ is surjective by the definition of $B(\mathcal{O}_1,\mathcal{O}_2)$ as the image of $P(\mathcal{O}_1,\mathcal{O}_2)$ under $p_2$.

Since $B(\mathcal{O}_1,\mathcal{O}_2)$ carries the trivial Poisson structure, the pullback $p_2^*f$ Poisson-commutes with every smooth function on $P^\circ$, for all $f\in C^\infty\bigl(B(\mathcal{O}_1,\mathcal{O}_2)\bigr)$. In particular, all Hamiltonian vector fields on $P^\circ$ are tangent to the fibers of $p_2|_{P^\circ}$.

For $b\in B(\mathcal{O}_1,\mathcal{O}_2)$, put $\mathcal{O}_b:=\chi^{-1}(b)\cap\mathfrak{g}^*_{\mathrm{reg}}$. The fiber of $p_2|_{P^\circ}$ is the corresponding open subset of
\begin{align*}
\left\{(x,y)\in\mathcal{O}_b\times\mathcal{O}_b^-:
\pi_1(x)\in\mathcal{O}_1,\ -\pi_2(y)\in\mathcal{O}_2\right\}\big/(K_1\times K_2),
\end{align*}
where $\mathcal{O}_b^-$ carries the opposite KKS form. The reduced symplectic form on this quotient is the nondegenerate form induced by the quotient Poisson tensor. Consequently, each connected component of the fiber is precisely a symplectic leaf.
%For completeness, the dimension of a fiber of $p_1$ includes the subgroup stabilizers. Fix $[g,\xi]$ and put $y:=\mathrm{Ad}_{g^{-1}}^*\xi$. Define \begin{align*}
%G_\xi&:=\{u\in G:\mathrm{Ad}_u^*\xi=\xi\},\qquad G_y:=\{u\in G:\mathrm{Ad}_u^*y=y\},\\
%H_1&:=K_1\cap G_\xi,\qquad H_2:=g(K_2\cap G_y)g^{-1}\subset G_\xi.
%\end{align*} Representatives of the same image class have the form \begin{align*}
%(k_1hgk_2^{-1},\mathrm{Ad}_{k_1}^*\xi),\qquad h\in G_\xi,
%\end{align*}
%and hence \begin{align*}
%p_1^{-1}(p_1([g,\xi]))\cong H_1\backslash G_\xi/H_2.
%\end{align*} Since $\xi$ is regular, $G_\xi$ is a maximal torus. The assumed free left-right action on $G$ implies $H_1\cap H_2=\{e\}$, so this fiber has dimension \begin{align*}
%\mathrm{rank}G-\dim H_1-\dim H_2.
%\end{align*} For $(K_1,K_2)=(\{e\},A)$ it is equivalently $G_y/A_y$, where $A_y:=A\cap G_y$, with dimension $\mathrm{rank}G-\dim A_y$.
\end{proof}

We treat cotangent bundle reduction in \cite{Kummer1981Reduction,MarsdenPerlmutter2000Orbit}. In what follows, let $\omega_{\mathrm{can}}$ be the canonical symplectic form on $T^*(K_1\backslash G/K_2)$ by abuse of notation. We now present the identification of $S(0,0)$ with $(T^*Q,\omega_{\mathrm{can}})$.
Recall that a differential form on a manifold with a Lie group action descends to the quotient only when it is both horizontal and invariant.
For background on quotient manifolds and basic forms, see \cite[Chapter 21] {Lee2013IntroductionSmoothManifolds}. We will use the definition below for checking basicness repeatedly throughout the rest of the paper.

\begin{definition}
\label{def:basic}
Let $K$ be a Lie group acting on the right on a smooth manifold $N$. For $\zeta\in\mathfrak k$, let $\widehat\zeta$ denote the fundamental vector field. A differential $\ell$-form $\alpha\in\Omega^\ell(N)$ is \emph{basic} if
\begin{align*}
\iota_{\widehat\zeta}\alpha=0\quad\text{and}\quad R_a^*\alpha=\alpha
\end{align*}
for every $\zeta\in\mathfrak k$ and every $a\in K$. Here $R_a$ denotes the right action by $a$. For connected $K$, invariance is equivalent to $\mathcal L_{\widehat\zeta}\alpha=0$ for all $\zeta\in\mathfrak k$. If the action is free and proper, every basic $\ell$-form is the pullback of a unique $\ell$-form on $N/K$.
\end{definition}

We now show that the zero-orbit leaf \begin{align*}
  S(0,0) := S(\{0\},\{0\})   = K_1\backslash \mu_{12}^{-1}(0,0)/K_2 %\label{eq:zeroleaf}
\end{align*} is canonically symplectomorphic to $T^*Q$, where $Q=K_1\backslash G/K_2$. Let $\pi: G\to Q$ be the quotient map. Since $K_1 \times K_2$ is a closed subgroup, $T_{\pi(g)}Q \cong T_gG/ \ker (d\pi)$. For each $g\in G$, the tangent map $d\pi: T_gG\to T_{\pi(g)}Q$ has a kernel \begin{align}
  \ker(d\pi)  = T_g(K_1 \cdot g)  +  T_g(g \cdot K_2) = \left\{X_1^L(g)  +  X_2^R(g)  :  \ X_1\in\mathfrak{k}_1,\ X_2\in\mathfrak{k}_2\right\}, \label{eq:kertan}
\end{align}  where $X_1^L(g)=(dR_g)_eX_1$ is the fundamental vector field for the left $K_1$-action and $X_2^R(g)=(dL_g)_eX_2$ is the fundamental vector field for the right $K_2$-action. In particular, at each $g\in G$ these are exactly the values of the corresponding right- and left-invariant vector fields, respectively.

\begin{proposition}
\label{prop:doubleidef}
    Let $G$ be a connected Lie group, and let $K_1,K_2\subset G$ be closed Lie subgroups. Assume that the left-right action of $K_1 \times K_2$ on $G$ is free and proper, such that $Q=K_1\backslash G/K_2$ is a smooth manifold. Let $\nu_i\in(\mathfrak{k}_i^*)^{K_i}$ and $\nu=(\nu_1,\nu_2)$. Let $\theta$ be a principal connection for the right action $g\cdot(k_1,k_2)=k_1^{-1}gk_2$, with the normalization and equivariance in Appendix \ref{app:twistedform}. Its curvature is $F_\theta=d\theta+\frac{1}{2}[\theta\wedge\theta]\in\Omega^2(G,\mathfrak{k}_1\oplus\mathfrak{k}_2)$. Define the closed form $\beta_\nu$ by $\pi^*\beta_\nu=\langle\nu,F_\theta\rangle$. Then there is a connection-dependent symplectomorphism \begin{align*}
\Phi:S(\{\nu_1\},\{\nu_2\})\longrightarrow (T^*Q,\omega_{\mathrm{can},Q}+\pi_Q^*\beta_\nu),
\end{align*}
where $\pi_Q:T^*Q\to Q$ is the bundle projection. In particular, for $\nu_1=\nu_2=0$, there is a canonical symplectomorphism   \begin{align*}
  \Phi:  \ S(0,0) \longrightarrow T^*Q   .
\end{align*}
\end{proposition} 
\begin{proof}
We keep the reduction value $\nu$ fixed and use $\xi\in\mathfrak{g}^*$ for the variable right-trivialized cotangent coordinate. We study the direct zero-level construction in Steps 1-4 and then extend it to arbitrary singleton orbits by a covector translation.

  \noindent\textbf{Step 1: identify the level set $\mu_{12}^{-1}(0,0)$ as an annihilator.}
Consider $(g,\xi)\in T^*G\cong G\times\mathfrak{g}^*$ and let $\alpha_g\in T_g^*G$ be the corresponding covector in \eqref{eq:right-trivialization}.
By \eqref{eq:coorlrmp}, the condition $\mu_{12}(g,\xi)=(0,0)$ is equivalent to \begin{align*}
  \pi_1(\xi)=0 \quad\text{and}\quad \pi_2\bigl(\mathrm{Ad}_{g^{-1}}^*\xi\bigr)=0,
\end{align*}
i.e., $\langle \xi,X_1\rangle=0$, and $\langle\xi,\mathrm{Ad}_g X_2\rangle=0$ with  $  X_i\in\mathfrak{k}_i$ for any $i = 1,2$.
On the other hand, the tangent spaces to the left and right orbits through $g$ are \begin{align*}
  T_g(K_1g) =\left\{\frac{d}{dt}\Big\vert_{t=0}\exp(tX_1)g\,:\,X_1\in\mathfrak{k}_1\right\},\quad  T_g(gK_2) =\left\{\frac{d}{dt}\Big\vert_{t=0}g\exp(tX_2)\,:\,X_2\in\mathfrak{k}_2\right\}.
\end{align*}
Applying $dR_{g^{-1}}$ gives $dR_{g^{-1}}\bigl(\frac{d}{dt}\vert_{t= 0}\exp(tX_1)g\bigr)=X_1$ and $dR_{g^{-1}}\bigl(\frac{d}{dt}\vert_{t= 0}g\exp(tX_2)\bigr)=\mathrm{Ad}_gX_2$. Hence, $\mu_{12}(g,\xi)=(0,0)$ if and only if $\alpha_g$ annihilates both subspaces $T_g(K_1g)$ and $T_g(gK_2)$.
Since $\pi:G\to Q=K_1\backslash G/K_2$ is the quotient map, its differential satisfies \begin{align*}
  \ker(d\pi_g)=T_g\bigl((K_1\times K_2)\cdot g\bigr)=T_g(K_1g)+T_g(gK_2).
\end{align*} Therefore, $ \mu_{12}^{-1}(0,0)=\mathrm{Ann}(\ker d\pi)\subset T^*G$. By the universal property of quotient vector spaces, there is a unique covector $\alpha_{\pi(g)}\in T^*_{\pi(g)}Q$ such that
\begin{align*}
  \alpha_{\pi(g)}(d\pi_g v)=\alpha_g(v)=\langle\xi,(dR_{g^{-1}})v\rangle
\end{align*} for all $v\in T_gG$.
%Consider the linear functional  \begin{align*}
  %\tilde\alpha_g :  T_gG \longrightarrow \mathbb{R},\qquad   \tilde\alpha_g(v) := \langle \xi,(dR_{g^{-1}})v\rangle.
%\end{align*}  Thus, the covector  $\widetilde\alpha_g(v) =\langle\xi,(dR_{g^{-1}})v\rangle$ with $v\in T_gG$, annihilates $\ker(d\pi_g)$.
%This map is $K_1\times K_2$-invariant:  for $(K_1,K_2)\in K_1\times K_2$, we have \begin{align*}
%  (g, \xi)\cdot (K_1,K_2) := \bigl(\mathrm{Ad}^*_{K_1}x,\ K_1gK_2^{-1}\bigr),
%\end{align*} and since $\pi(K_1gK_2^{-1})=\pi(g)$ and $(g, \xi)$ and $(\mathrm{Ad}^*_{K_1}x,K_1gK_2^{-1})$ determine the same functional on $T_{\pi(g)}Q$ via the formula above, the associated covector $\alpha_{\pi(g)}$ does not depend on the choice of representative in the $K_1\times K_2$-orbit. Thus, the map descends to a bijection \begin{align*}
 %\Phi:  S(0,0)=K_1\backslash\mu_{12}^{-1}(0,0)/K_2  \longrightarrow T^*M.
%\end{align*}

 \noindent\textbf{Step 2: Well-definedness, smoothness, and invariance of $\Phi_0$.} Fix $(g,\xi)\in\mu_{12}^{-1}(0,0)$ and write $q:=\pi(g)\in Q$. For $w\in T_qQ$, take any $v\in T_gG$ with $d\pi_g(v)=w$ (possible since $d\pi_g$ is surjective) and define
\begin{align}
\alpha_q(w):=\alpha_g(v)=\langle\xi,(dR_{g^{-1}})_gv\rangle. \label{eq:alphaqdef}
\end{align}
This is well defined: if $v'$ is another lift of $w$, then $v'-v\in\ker(d\pi_g)$ and Step 1 give $\alpha_g(v'-v)=0$. Hence, $\alpha_g(v')=\alpha_g(v)$. Thus, $\alpha_g=(d\pi_g)^*\alpha_q$ and the assignment
\begin{align*}
\Phi_0:\mu_{12}^{-1}(0,0)\longrightarrow T^*Q,\qquad \Phi_0(g,\xi)=(q,\alpha_q),
\end{align*}
is smooth.

To see $K_1\times K_2$-invariance, let $(k_1,k_2)\in K_1\times K_2$ and put $g':=k_1gk_2^{-1}$. In right trivialization, the cotangent-lifted action is $(g,\xi)\mapsto(g',\xi')$ with $\xi'=\mathrm{Ad}_{k_1}^*\xi$. Since $\pi(g')=\pi(g)$, we have $q'\!=\!q$. For $w\in T_qQ$ and any $v\in T_gG$ with $d\pi_g(v)=w$, set $v':=(dL_{k_1})_{gk_2^{-1}}(dR_{k_2^{-1}})_gv\in T_{g'}G$, so that $d\pi_{g'}(v')=w$. Then
\begin{align*}
\alpha_{g'}(v')
&=\langle\xi',(dR_{(g')^{-1}})_{g'}v'\rangle
=\langle\mathrm{Ad}_{k_1}^*\xi,\,\mathrm{Ad}_{k_1}(dR_{g^{-1}})_gv\rangle
=\langle\xi,(dR_{g^{-1}})_gv\rangle
=\alpha_g(v).
\end{align*}
Hence the covector $\alpha_q$ defined by \eqref{eq:alphaqdef} is unchanged along $(K_1\times K_2)$-orbits, so $\Phi_0$ descends to a smooth map
\begin{align*}
\Phi:S(0,0)=K_1\backslash\mu_{12}^{-1}(0,0)/K_2\longrightarrow T^*Q.
\end{align*}

\noindent\textbf{Step 3: bijectivity of $\Phi$.}
 We first prove surjectivity. Let $q\in Q$ and take any $g\in G$ with $\pi(g)=q$. Let $\alpha_q\in T_q^*Q$ be arbitrary.
Define the unique covector $\xi\in\mathfrak{g}^*$ by
\begin{align*}
  \xi(u) := \alpha_q\bigl(d\pi_g((dR_g)_e u)\bigr),\qquad u\in\mathfrak{g}.
\end{align*}
Then, by the definition of the right trivialization, we have
\begin{align*}
  \langle \xi,(dR_{g^{-1}})v\rangle   = \alpha_q\bigl(d\pi_g(v)\bigr), \qquad v\in T_gG.
\end{align*}
Since $\alpha_q\circ d\pi_g$ vanishes on $\ker(d\pi_g)$, the above indeed produces a well-defined linear functional on $\mathfrak{g}$.
%hence an element $X \in \mathfrak{g}  $. 
By construction, $\alpha_g$ annihilates $\ker(d\pi)$, and hence $\mu_{12}(g, \xi)=(0,0)$, so $(g, \xi)\in\mu_{12}^{-1}(0,0)$ and $\Phi_0(g, \xi)=(q,\alpha_q)$.
Thus, $\Phi$ is surjective.

For injectivity, suppose that $\Phi\bigl([g,\xi]\bigr)=\Phi\bigl([g',\xi']\bigr)$. Choose representatives $(g,\xi)$ and $(g',\xi')$ (in $\mu_{12}^{-1}(0,0)$) with the same image $(q,\alpha_q)\in T^*Q$. By the uniqueness in Step 1, this implies that $g$ and $g'$ represent the same point $q\in Q$ and that the corresponding covectors are $(d\pi_g)^*\alpha_q$ and $(d\pi_{g'})^*\alpha_q$, respectively. Therefore, $(g,\xi)$ and $(g',\xi')$ lie in the same $K_1\times K_2$-orbit, so $[g,\xi]=[g',\xi']$.
Thus, $\Phi$ is injective. The same local construction gives a smooth inverse, so $\Phi$ is a diffeomorphism.
%The above argument also shows that $\Phi$ has a smooth inverse, constructed fiberwise by choosing a representative $g$ of $q$ and defining $x$ from $\alpha_q$ as in the surjectivity part. Therefore, $\Phi$ is a diffeomorphism.

\noindent\textbf{Step 4:  $\Phi$ is symplectic.}
Let $\vartheta_{G}$ and $\vartheta_{Q}$ denote the tautological one-forms on $T^*G$ and $T^*Q$, respectively. Write $\pi_G:T^*G\to G$, $\pi_G(g,\xi)=g$, to distinguish the cotangent projection from $\pi:G\to Q$. Recall that at a point $(g,\xi)\in T^*G$ with $(\dot{g},\dot{\xi}) \in T_{(g,\xi)}T^*G$, the canonical $1$-form is defined by
\begin{align*}
  \vartheta_{G}(g,\xi)(\dot{g},\dot\xi)  := \alpha_g\bigl(d\pi_G(\dot g,\dot\xi)\bigr)=\alpha_g(\dot g)  = \big\langle\xi,(dR_{g^{-1}})\dot{g}\big\rangle,
\end{align*} and at $(q,\alpha_q)\in T^*Q$, we have $\vartheta_{Q}(q,\alpha_q)(\dot q,\dot\alpha_q):=\alpha_q(\dot q)$.  Consider the inclusion $i: \mu_{12}^{-1}(0,0)\hookrightarrow T^*G$ and the quotient map $p: \mu_{12}^{-1}(0,0)\to S(0,0)$. By construction, $p^*\Phi^*\vartheta_{Q}=\Phi_0^*\vartheta_{Q}$. Using the definition of $\Phi_0$, we compute for $(\dot{g},\dot{\xi})\in T_{(g,\xi)}\mu_{12}^{-1}(0,0)$: \begin{align*}
  \Phi_0^*\vartheta_{Q}(g,\xi)(\dot{g},\dot{\xi})  &= \vartheta_{Q}\bigl(\Phi_0(g,\xi)\bigr)\bigl(d\Phi_0(\dot{g},\dot{\xi})\bigr)% \\
 % &= \alpha_q\bigl(d\pi(\dot{g})\bigr)\\
  %&= \big\langle \xi,(dR_{g^{-1}})\dot{g}\big\rangle\\
   = \vartheta_{G}(g,\xi)(\dot{g},\dot{\xi}).
\end{align*} Thus,  $i^*\vartheta_{G} = \Phi_0^*\vartheta_{Q} = p^*\Phi^*\vartheta_{Q}$, where $p:\mu_{12}^{-1}(0,0)\to S(0,0)$ is the quotient map.
%Now pass to the symplectic forms by taking exterior derivatives.
Since $\omega_{\mathrm{can}} = -d\vartheta_{G}$ on $T^*G$ and $\omega_{T^*Q}= - d\vartheta_{Q}$ on $T^*Q$, we obtain \begin{align*}
  p^*\omega_{\mathrm{red}} = i^*\omega_{\mathrm{can}}   =  % i^*(d\vartheta_{G})  = 
  - d(i^*\vartheta_{G}) =  p^*\Phi^*\bigl(-d\vartheta_{Q}\bigr) = p^*\Phi^*\omega_{T^*Q}.
\end{align*} As $p$ is a surjective submersion, this implies $\omega_{\mathrm{red}}=\Phi^*\omega_{T^*Q}$, i.e., $\Phi$ is a symplectomorphism.

\noindent\textbf{Step 5: nonzero singleton orbits.}
Put $\tau:=\langle\nu,\theta\rangle\in\Omega^1(G)$. For $X=(X_1,X_2)\in\mathfrak k_1\oplus\mathfrak k_2$, the fundamental vector field of the principal action is $\widehat X(g)=-(dR_g)_eX_1+(dL_g)_eX_2$, so $\tau(\widehat X)=\langle\nu,X\rangle$. The Hamiltonian action is generated by $-\widehat X$, so $\alpha_g(\widehat X(g)) = -\langle\mu_{12}(g,\xi),X\rangle$. Hence, for $(g,\xi)\in\mu_{12}^{-1}(\nu)$, \begin{align*}
(\alpha_g+\tau_g)\bigl(\widehat X(g)\bigr)=-\langle\nu,X\rangle+\langle\nu,X\rangle=0,
\end{align*}
that is, $\alpha_g+\tau_g$ annihilates $\ker d\pi_g$. Since $\nu$ is $\mathrm{Ad}^*(K_1\times K_2)$-invariant, $\tau$ is $K_1\times K_2$-invariant, and the fiber translation $t_\tau\colon\alpha_g\mapsto\alpha_g+\tau_g$ restricts to an equivariant diffeomorphism $\mu_{12}^{-1}(\nu)\to\mu_{12}^{-1}(0)$. We define $\Phi$ as the map induced by $\Phi_0\circ t_\tau$.

Let $i$ and $p$ be the inclusion of $\mu_{12}^{-1}(\nu)$ into $T^*G$ and the quotient map onto $S(\{\nu_1\},\{\nu_2\})$. Since $t_\tau^*\vartheta_G=\vartheta_G+\pi_G^*\tau$, Step 4 gives \begin{align*}
p^*\Phi^*\vartheta_Q=i^*\vartheta_G+\bigl(\pi_G|_{\mu_{12}^{-1}(\nu)}\bigr)^*\tau.
\end{align*}
As $\nu$ vanishes on $[\mathfrak k_1\oplus\mathfrak k_2,\mathfrak k_1\oplus\mathfrak k_2]$, we have $d\tau=\langle\nu,F_\theta\rangle=\pi^*\beta_\nu$. Applying $-d$ and using $\pi_Q\circ\Phi\circ p=\pi\circ\pi_G$ on $\mu_{12}^{-1}(\nu)$, we obtain \begin{align*}
p^*\Phi^*\bigl(\omega_{\mathrm{can},Q}+\pi_Q^*\beta_\nu\bigr)=i^*\omega_{\mathrm{can}}=p^*\omega_{\mathrm{red}},
\end{align*}
and the injectivity of $p^*$ gives $\Phi^*\bigl(\omega_{\mathrm{can},Q}+\pi_Q^*\beta_\nu\bigr)=\omega_{\mathrm{red}}$.
\end{proof}

% The discussion of singleton (point) coadjoint orbits is collected in Remark \ref{rem:spinandtwist} below.

\begin{corollary}
\label{coro:superintetr}
   Let $G$ be a connected Lie group, and let $K_1,K_2 \subset G$ be closed subgroups for which the left-right action on $G$ is free and proper, such that $Q = K_1 \backslash G /K_2$ is a smooth manifold.
Suppose that $\mathcal{O}_1 \subset \mathfrak{k}_1^*$, and $\mathcal{O}_2 \subset \mathfrak{k}_2^*$ are coadjoint orbits such that $S(\mathcal{O}_1,\mathcal{O}_2) \cong T^*Q$. Then the coadjoint orbits associated with the symplectic leaf must satisfy $\dim \mathcal{O}_1 = \dim \mathcal{O}_2 = 0$. In particular, if $K_1$ and $K_2$ are connected, then $\mathcal{O}_1$ and $\mathcal{O}_2 $ are singleton coadjoint orbits.
\end{corollary} 
\begin{proof}
    Since $\dim Q = \dim G -\dim K_1 - \dim K_2$, we have $ \dim T^*Q = 2 \dim G - 2 \dim K_1 - 2 \dim K_2$.
On the other hand, since $\dim S(\mathcal{O}_1,\mathcal{O}_2) = 2 \dim G - 2 \dim K_1 - 2\dim K_2 + \dim \mathcal{O}_1 + \dim \mathcal{O}_2$.
Comparing two-dimensional formulas yields $\dim  \mathcal{O}_1 + \dim \mathcal{O}_2 = 0$. Since $\dim \mathcal{O}_i \geq 0$, then $\dim  \mathcal{O}_1 = \dim \mathcal{O}_2 = 0$.
\end{proof}

\begin{remark}\label{rem:spinandtwist}
A coadjoint orbit $\mathcal{O} = \{\nu\} \subset \mathfrak{k}^*$ is a singleton if and only if $\nu \in (\mathfrak{k}^*)^K$. While reduction at zero orbits yields the canonical cotangent bundle $T^*Q$ (Proposition \ref{prop:doubleidef}), reduction at a nonzero singleton coadjoint orbit $(\nu_1, \nu_2)$ yields a magnetic cotangent bundle $(T^*Q, \omega_{\mathrm{can},Q} + \pi_Q^* \beta_\nu)$, where $\pi^* \beta_\nu = \langle \nu, F_\theta \rangle$ for any principal connection $\theta$.
\end{remark}

\subsection{Specialization to the centralizer \texorpdfstring{$A=Z_G(\gamma)$}{A = ZG(gamma)} and the orbits \texorpdfstring{$(\mathcal{O}_1,\mathcal{O}_2)$}{(O1,O2)}}
\label{sec:specialization}

In what follows, we use the covector $\lambda=B^\flat(\gamma) \in \mathfrak{g}^*$ and its restriction $\pi_{\mathfrak{a}}(\lambda)\in\mathfrak{a}^*$. We take \begin{align*}
K_1=\{e\},\qquad \mathcal{O}_1= \{0\}\subseteq\mathfrak{k}_1^*=\{0\}, \qquad K_2=A,\qquad \mathcal{O}_2=\{\varepsilon \pi_{\mathfrak{a}}(\lambda) \}\subseteq\mathfrak{a}^*.
\end{align*}
This is well-defined as $A$ centralizes $\gamma$. Hence,  $\lambda$ is $\mathrm{Ad}^*(A)$-invariant, and therefore   $\pi_{\mathfrak{a}}(\lambda)$ is a fixed coadjoint covector. In particular, $\mathcal{O}_2$ is a singleton coadjoint orbit in $\mathfrak{a}^*$. By the definition of the left and right moment maps in \eqref{eq:coorlrmp}, let $\pi_\mathfrak{a}: \mathfrak{g}^* \rightarrow \mathfrak{a}^*$. We have: \begin{align*}
    \mu_A^R: T^*G \cong  G \times \mathfrak{g}^* \longrightarrow \mathfrak{a}^*, \qquad \mu_A^R(g, \xi):= -\pi_\mathfrak{a}(\mathrm{Ad}_{g^{-1}}^*\xi)
\end{align*} so that, for any $\zeta \in \mathfrak{a}$, \begin{align*}
    \left\langle \mu_A^R(g,\xi),\zeta \right\rangle = -\langle \mathrm{Ad}_{g^{-1}}^*\xi,\zeta\rangle = - \langle \xi, \mathrm{Ad}_g\zeta \rangle.
\end{align*}  With these choices, the right constraint $\mu_A^R(g, \xi)\in \mathcal{O}_2$ reads \begin{align*}
-\pi_{\mathfrak{a}} \bigl(\mathrm{Ad}_{g^{-1}}^*\xi\bigr) = \varepsilon \pi_{\mathfrak{a}}(\lambda)  \qquad \Longleftrightarrow \qquad \pi_{\mathfrak{a}} \bigl(\mathrm{Ad}_{g^{-1}}^*\xi\bigr)= -\varepsilon \pi_{\mathfrak{a}}(\lambda) , %\label{eq:rightmomentmap}
\end{align*} i.e.,  writing $y=\mathrm{Ad}_{g^{-1}}^*\xi$, the constraint is $\pi_{\mathfrak{a}}(y+\varepsilon\lambda)=0$.
%This is exactly the affine shift appearing in \eqref{eq:chainad}.

In the rest of this subsection, let $M:=G/A$. Fix the homogeneous principal $A$-connection \begin{align*}
\theta_g:=\mathrm{pr}_{\mathfrak{a}}\circ(dL_{g^{-1}})_g\in\Omega^1(G,\mathfrak{a}),
\end{align*}
where $\mathrm{pr}_{\mathfrak{a}}:\mathfrak{g}\to\mathfrak{a}$ is the projection along $\mathfrak{m}$ for the fixed reductive decomposition $\mathfrak{g}=\mathfrak{a}\oplus\mathfrak{m}$. (This is different from the restriction map $\pi_{\mathfrak{a}}:\mathfrak{g}^*\to\mathfrak{a}^*$ used in the moment map formulas.)
It is normalized for the principal right action $g\cdot a=ga$:
\begin{align*}
\theta_g((dL_g)_e\zeta)=\zeta,\qquad R_a^*\theta=\mathrm{Ad}_{a^{-1}}\theta,
\end{align*}
and its curvature is $F_\theta=d\theta+\frac{1}{2}[\theta\wedge\theta]\in\Omega^2(G,\mathfrak{a})$. For the Hamiltonian inverse-right action $a\cdot g=ga^{-1}$, the fundamental vector is $-(dL_g)_e\zeta$, and hence it has a $\theta$-value of $-\zeta$.

Since $\gamma\in \mathfrak{a}$ is $\mathrm{Ad}(A)$-fixed, the $\mathbb{R}$-valued $2$-form \begin{align*}
\widetilde{\sigma}_{\gamma}:=\langle\lambda,F_\theta\rangle\in\Omega^2(G)
\end{align*}
is basic and descends to a closed $2$-form $\sigma_{\gamma}\in\Omega^2(M)$ (Appendix \ref{app:twistedform}, Proposition \ref{prop:curvature}). With this homogeneous choice of $\theta$, the descended form is $\sigma_{\gamma}$: for $X,Y\in\mathfrak m$, at $e$ we have $F_\theta(X,Y)=-\mathrm{pr}_{\mathfrak a}[X,Y]$, so $\langle\lambda,F_\theta\rangle(X,Y)=-B(\gamma,[X,Y])$, which is $\rho^*\sigma_{\gamma}(X,Y)$. Both sides are left-invariant and horizontal. We then define the twisted form \begin{align}
\omega_\varepsilon:=\omega_{\mathrm{can}} + \varepsilon\,\pi^*\sigma_{\gamma}  , \label{eq:twistedsymplec}
\end{align} on  $T^*M$ (see, Theorem \ref{thm:superin}), where $\pi:T^*M\to M$ is the bundle projection. By Theorem \ref{thm:twistedcota}, the form  \eqref{eq:twistedsymplec} is symplectic.

\begin{proposition}
\label{prop:mincoup}
Let $A=Z_G(\gamma)$, $M=G/A$, $(K_1,K_2)=(\{e\},A)$, and $(\mathcal{O}_1,\mathcal{O}_2)=(\{0\},\{\varepsilon\pi_{\mathfrak{a}}(\lambda)\})$. Let $\rho:G\to G/A$ be the quotient map for the principal right $A$-action $g\cdot a=ga$. (The Hamiltonian convention is the left action $a\cdot g=ga^{-1}$.) Let $\theta \in \Omega^1(G,\mathfrak{a})$ be the homogeneous principal $A$-connection fixed above, and let $F_\theta \in \Omega^2(G,\mathfrak{a})$ be its curvature form. Put $C_\varepsilon := (\mu_A^R)^{-1}(\varepsilon \pi_{\mathfrak{a}}(\lambda))$. Then %with parameter $\eta=-\varepsilon \gamma$, 
\begin{align*}
    S(\mathcal{O}_1,\mathcal{O}_2) = \mu_{12}^{-1}(0,\varepsilon \pi_{\mathfrak{a}}(\lambda) )/(\{e\}\times A) = (\mu_A^R)^{-1}(\varepsilon \pi_{\mathfrak{a}}(\lambda))/A = C_\varepsilon/A
\end{align*} is symplectomorphic to $\bigl(T^*(G/A),\,\omega_{\mathrm{can}} + \varepsilon\,\pi^*\sigma_{\gamma}\bigr)$.
%In particular, for $\varepsilon=0$ this recovers Proposition 2.10 with $Q=G/A$.
\end{proposition}

\begin{proof}
This follows by specializing Proposition \ref{prop:doubleidef} to $K_1=\{e\}$, $K_2=A$, $\nu_1=0$, and $\nu_2=\varepsilon\pi_{\mathfrak{a}}(\lambda)$. Under these choices, $Q=G/A=M$ and the curvature pairing satisfies \begin{align*}
\rho^*\beta_{(0,\varepsilon\pi_{\mathfrak{a}}(\lambda))}=\varepsilon\langle\lambda,F_\theta\rangle=\varepsilon\,\rho^*\sigma_{\gamma},
\end{align*}
so the reduced form is $\omega_{\mathrm{can}}+\varepsilon\pi^*\sigma_{\gamma}$. The map $\Phi_S:C_\varepsilon/A\to T^*M$ is a symplectomorphism by Proposition \ref{prop:doubleidef}, and is given explicitly by
\begin{align*}
\Phi_S([g,\xi]_A)=(gA,\alpha_{gA}),\qquad (d\rho_g)^*\alpha_{gA}=\alpha_g+\varepsilon\langle\lambda,\theta_g\rangle,
\end{align*}
where $\alpha_g$ is the covector in \eqref{eq:right-trivialization}.
\end{proof}

\begin{remark}\label{rem:homogeneouscoord}
For the $G$-invariant connection associated with the reductive decomposition $\mathfrak{g}=\mathfrak{a}\oplus\mathfrak{m}$, we identify $T^*M\cong G\times_A\widetilde{\mathfrak{a}}$. In terms of coordinates, the diffeomorphism $\Phi_S:C_\varepsilon/A\to T^*M$ is given by \begin{align*}
\Phi_S([g,\xi]_A) &= [g,\eta]_A, \qquad \text{where }\ \mathrm{Ad}_{g^{-1}}^*\xi = \eta-\varepsilon \lambda \ \text{with }\eta\in \widetilde{\mathfrak{a}} ,
\end{align*}
(i.e., $\eta=\mathrm{Ad}_{g^{-1}}^*\xi+\varepsilon\lambda$ and the constraint defining $C_\varepsilon$ fixes the $\mathfrak{a}^*$-component to be $-\varepsilon\pi_{\mathfrak{a}}(\lambda)$). In other words, $\eta$ is the unique element of $\widetilde{\mathfrak{a}}$ whose associated covector $\alpha_{gA}\in T^*_{gA}M$   satisfies $(d\rho_g)^*\alpha_{gA}=\alpha_g+\varepsilon\langle\lambda,\theta_g\rangle$. Its inverse  $\Psi_S:=\Phi_S^{-1}:T^*M\cong G\times_A\widetilde{\mathfrak{a}}\to C_\varepsilon/A$ is \begin{align*}
\Psi_S([g,\eta]_A) &= [g,\mathrm{Ad}_g^*(\eta-\varepsilon \lambda )]_A.
\end{align*}
%The diffeomorphism $\Phi_S$ depends on the choice of connection $\theta$, but changing $\theta$ changes $\Phi_S$ by a fiber translation on $T^*M$ (hence by a canonical symplectomorphism). Thus, once the homogeneous connection is fixed, the displayed maps are mutually inverse symplectomorphisms.
\end{remark}

We now explain the intermediate space and base space under the phase space $S(\mathcal{O}_1,\mathcal{O}_2)$ in a way that makes the comparison with $P(\mathcal{O}_1,\mathcal{O}_2)$ and $B(\mathcal{O}_1,\mathcal{O}_2)$ tautological.
Let $\chi:\mathfrak{g}^*\to\mathfrak{g}^*//G$ denote the coadjoint quotient map introduced in \eqref{eq:affinegit}.
Recall that, under the magnetic phase space $T^*(G/A)$, the base space is \begin{align*}
\mathcal{B} =   \overline{\chi\left((\widetilde{\mathfrak{a}}-\varepsilon\lambda) \cap \mathfrak{g}_\mathrm{reg}^*\right)}  \ \subseteq\ \mathfrak{g}^* // G,
\end{align*} where $\overline{\cdot}$ means the topological closure.
Let \begin{align*}
R_0   \cong  \mathbb{R}[\mathcal{B} ] \cong \mathrm{Im}\bigl(S(\mathfrak{g})^G\longrightarrow   \mathbb{R}[\widetilde{\mathfrak{a}} -\varepsilon  \lambda] \bigr),
\end{align*} where $\mathbb{R}[\mathcal B]$ means restrictions of invariant polynomials and the map is restricted along the inclusion $\widetilde{\mathfrak{a}} - \varepsilon\lambda\hookrightarrow \mathfrak{g}^*$.
Since $\widetilde{\mathfrak{a}}-\varepsilon\lambda$ is $\mathrm{Ad}^*(A)$ -stable, its affine invariant-theory quotient is $\mathrm{Spec}\,\mathbb{R}[\widetilde{\mathfrak{a}}-\varepsilon\lambda]^A$. In the geometric chains, $(\widetilde{\mathfrak{a}}-\varepsilon\lambda)//A$ is the affine GIT quotient, restricted to the chosen smooth piece.
The geometric chains use the smooth spaces $U_0,U_1$ fixed in \eqref{eq:comparison-domains}.
The restriction map   $S(\mathfrak{g})^G \rightarrow \mathbb{R}[\widetilde{\mathfrak{a}}-\varepsilon\lambda]^A$   factors through $R_0$ and induces a canonical morphism \begin{align*}
  \mathrm{Spec}\,\mathbb{R}[\widetilde{\mathfrak{a}} -\varepsilon \lambda]^A \ \longrightarrow\ \mathrm{Spec}\  R_0.
\end{align*}
On the other hand, the intermediate fiber product is
\begin{align}
\mathrm{Ad}^* (G)( \widetilde{\mathfrak{a}} -\varepsilon \lambda)\times _{ \mathrm{Spec} R_0}( \widetilde{\mathfrak{a}} -\varepsilon \lambda )// A . \label{eq:fibermide}
\end{align}
On the subset $U_1$ used below, a point of \eqref{eq:fibermide} is represented by a pair $(x,[y]_A)$ with
\begin{align*}
y\in V,\qquad x\in\mathfrak{g}^*_{\mathrm{reg}},\qquad \chi(x)=\chi(y),
\end{align*}
where $V/A$ is identified with its image in the real points of the affine GIT quotient via the evaluation of invariant polynomials.

In the comparison theorem below, we use \eqref{eq:fibermide} only on the smooth subset $U_1$, with $y\in V$. No effective freeness of the $A$-action on this slice is assumed. Let $\{\cdot,\cdot\}_{\mathrm{fib}}$ be the Poisson bracket on $U_1$ given by Theorem \ref{thm:superin}. Since
\begin{align*}
\Pi_1:(T^*M,\omega_\varepsilon)\big\vert_U\longrightarrow  U_1
\end{align*}
is a surjective Poisson submersion, the bracket $\{\cdot,\cdot\}_{\mathrm{fib}}$ is determined by
\begin{align}
\Pi_1^*\{f_1,f_2\}_{\mathrm{fib}}=\{\Pi_1^*f_1,\Pi_1^*f_2\}_{T^*M}\qquad \text{for all }f_1,f_2\in C^\infty(U_1), \label{eq:definefibbracket}
\end{align}
where $\{\cdot,\cdot\}_{T^*M}$ is the Poisson bracket associated with $\omega_\varepsilon$.
%Equivalently (via the identification with the principal locus of $P(\mathcal{O}_1,\mathcal{O}_2)$ established below), this Poisson structure is the quotient Poisson structure induced from the product Lie-Poisson structure on $\mathfrak{g}^*\times \mathfrak{g}^*$, with the second factor carrying the opposite bracket. 
On  $\mathfrak{g}_\mathrm{reg}^*$  (see \eqref{eq:regular}), the quotient map $\chi$ has maximal rank, and its fibers are single $G$-orbits of constant type.
We retain exactly the orbit-type and constant rank restrictions used to construct $U,U_1,U_0$ before Theorem \ref{thm:superin}. Thus, the intermediate fiber product is used only through $U_1$, and the base only through its embedded image $U_0$.
We now provide the equivalence between these two chains:

Fix $\gamma\in\mathfrak{g}$, put $A:=Z_G(\gamma)$, and take $(K_1,K_2)=(\{e\},A)$ with $\mathcal{O}_1=\{0\}$ and $\mathcal{O}_2=\{\varepsilon\pi_{\mathfrak a}(\lambda)\}\subset\mathfrak a^*$. Let $U,U_1,U_0$ be as in \eqref{eq:comparison-domains}, and set
\begin{align*}
S_U(\mathcal{O}_1,\mathcal{O}_2)&:=\Phi_S^{-1}(U)=\bigl\{[g,\xi]_A\in S(\mathcal{O}_1,\mathcal{O}_2):\ \mathrm{Ad}_{g^{-1}}^*\xi\in V\bigr\},\\
P_U(\mathcal{O}_1,\mathcal{O}_2)&:=p_1\bigl(S_U(\mathcal{O}_1,\mathcal{O}_2)\bigr),\qquad
B_U(\mathcal{O}_1,\mathcal{O}_2):=p_2\bigl(P_U(\mathcal{O}_1,\mathcal{O}_2)\bigr).
\end{align*}

\begin{theorem}
\label{thm:specequiv}
Let $G$ be compact, connected, and semisimple, $\gamma\in\mathfrak{g}$, and $A:=Z_G(\gamma)$, and assume \eqref{eq:nonempreg}. Then the diagram below commutes, and its vertical arrows are Poisson diffeomorphisms. In particular, the two chains are equivalent in the sense of Definition \ref{def:equidefi}.
\begin{center}
     $\begin{tikzcd}
(T^*M,\omega_\varepsilon)\big\vert_U \arrow[r,"\Pi_1"] \arrow[d,"\Phi_S^{-1}"'] &
U_1 \arrow[r,"\Pi_2"] \arrow[d,"\Phi_P"'] &
U_0 \arrow[d,"\Phi_B"'] \\
S_U(\mathcal{O}_1,\mathcal{O}_2) \arrow[r,"p_1"'] &
P_U(\mathcal{O}_1,\mathcal{O}_2) \arrow[r,"p_2"'] &
B_U(\mathcal{O}_1,\mathcal{O}_2).
\end{tikzcd}$
 \end{center}
\end{theorem}

\begin{proof}
 First, Proposition \ref{prop:mincoup} gives a symplectomorphism
\begin{align*}
\Phi_{S}\vert_{S_U(\mathcal{O}_1,\mathcal{O}_2)}: S_U(\mathcal{O}_1,\mathcal{O}_2) \xrightarrow{\sim} (T^*M,\omega_{\varepsilon})\vert_U.
\end{align*}
Hence, $\Phi_{S}$ is a Poisson diffeomorphism.

Next, we construct the middle identification. Write $\mathcal{O}_G(x):=G\cdot x$ for the coadjoint orbit.
 By definition, \begin{align*}
P_U(\mathcal{O}_1, \mathcal{O}_2) = \{(x,y) \in \mathfrak{g}^*_{\mathrm{reg}} \times V :  \pi_1(x) \in \mathcal{O}_1,\ -\pi_2(y) \in \mathcal{O}_2,\ \mathcal{O}_G (x)=\mathcal{O}_G (y)\}/(e \times A).
\end{align*} Since $K_1=e$, the condition $\pi_1(x) \in \mathcal{O}_1$ is automatic.
Since $\mathcal{O}_2=\{\varepsilon\pi_{\mathfrak{a}}(\lambda)\}$, the condition $-\pi_2(y) \in \mathcal{O}_2$ is equivalent to $y \in \widetilde{\mathfrak{a}}-\varepsilon\lambda$.
Therefore, every class $[(x,y)] \in P_U(\mathcal{O}_1, \mathcal{O}_2)$ satisfies $y \in \widetilde{\mathfrak{a}}-\varepsilon\lambda$ and $x \in \mathcal{O}_G (y) \subset \mathrm{Ad}^*(G)(\widetilde{\mathfrak{a}}-\varepsilon\lambda)$. On $\mathfrak{g}_\mathrm{reg}^*$, the coadjoint quotient separates $G$-orbits of constant type, so $\chi(x) = \chi(y)$ if and only if $\mathcal{O}_G (x)=\mathcal{O}_G (y)$. Hence, after restricting to $y\in V$ as in $U_1$, $P_U(\mathcal{O}_1, \mathcal{O}_2)$ is canonically identified with the subset $U_1$ of the fiber product
\begin{align*}
\mathrm{Ad}^* (G)( \widetilde{\mathfrak{a}} -\varepsilon \lambda ) \times_{\mathrm{Spec}\,R_0} ( \widetilde{\mathfrak{a}} -\varepsilon \lambda )//A.
\end{align*}
Define
\begin{align*}
\Phi_P(x,[y]_{A}) := [(x,y)] \in P_U(\mathcal{O}_1, \mathcal{O}_2).
\end{align*}
This is well-defined since replacing $y$ with $a\cdot y$ changes $(x,y)$ by the $\{e\}\times A$-action. Its inverse is $\Psi_P([(x,y)]) := (x,[y]_{A})$. Both maps are induced by the identity on the same constrained pair set, with $A$ acting on the second factor, hence are smooth inverse maps. Thus, $\Phi_P$ is a diffeomorphism on $U_1$. We now prove that $\Phi_P$ is Poisson. Let $\{\cdot,\cdot\}_{\mathrm{fib}}$ denote the given Poisson bracket on $U_1$, and let $\{\cdot,\cdot\}_{P_U}$ be the Poisson bracket on $P_U(\mathcal{O}_1, \mathcal{O}_2)$. Let $\{\cdot,\cdot\}_{\mathrm{tr}}$ be the bracket on $U_1$ transported from $P_U(\mathcal{O}_1,\mathcal{O}_2)$ by the diffeomorphism $\Phi_P$, namely \begin{align*}
\{f_1,f_2\}_{\mathrm{tr}} :=\Phi_P^*\left(\{(\Phi_P^{-1})^*f_1,(\Phi_P^{-1})^*f_2\}_{P_U}\right).
\end{align*}
 The coordinate formulas give $\Phi_P\circ\Pi_1=p_1\circ\Phi_S^{-1}$: both sides send $[g,X]_A$ to $[(\mathrm{Ad}_g^*(X-\varepsilon\lambda),X-\varepsilon\lambda)]$. Since $p_1$ and $\Phi_S^{-1}$ are Poisson, $\Pi_1$ is Poisson. Hence, for all $f_1,f_2\in C^\infty(U_1)$, \begin{align*}
\Pi_1^*\{f_1,f_2\}_{\mathrm{tr}} =\{\Pi_1^*f_1,\Pi_1^*f_2\}_{T^*M}.
\end{align*}
By Theorem \ref{thm:superin}, $\Pi_1$ is Poisson for the bracket $\{\cdot,\cdot\}_{\mathrm{fib}}$, so $\Pi_1^*\{f_1,f_2\}_{\mathrm{fib}} =\{ \Pi_1^*f_1, \Pi_1^*f_2\}_{T^*M}$. Since $\Pi_1$ is a surjective submersion on $U$, $\Pi_1^*$ is injective. Therefore, $\{f_1,f_2\}_{\mathrm{tr}}=\{f_1,f_2\}_{\mathrm{fib}}$. Thus, the transported Poisson structure agrees with the given fiber-product Poisson structure, and $\Phi_P$ is a Poisson diffeomorphism.

We now identify the base spaces. Let $B(\mathcal{O}_1,\mathcal{O}_2)$ be the full space defined in \eqref{eq:base}. Then the base used here is its image restriction $B_U(\mathcal{O}_1,\mathcal{O}_2)$.
% \begin{align*}
% B_U(\mathcal{O}_1, \mathcal{O}_2) = \chi\bigl(\{Y \in \mathfrak{g} :  -\pi_2(Y) \in \mathcal{O}_2 \text{ and there exists } X \in \mathfrak{g} \text{ with } \pi_1(X) \in \mathcal{O}_1,\ \mathcal{O}_G (X)=\mathcal{O}_G (Y)\}\bigr).
% \end{align*}
By definition,
\begin{align*}
B_U(\mathcal{O}_1,\mathcal{O}_2) = p_2\bigl(P_U(\mathcal{O}_1,\mathcal{O}_2)\bigr)  = p_2\bigl(p_1(S_U(\mathcal{O}_1,\mathcal{O}_2))\bigr).
\end{align*}
On the other hand, on the subset under consideration, we have $y\in V$ and $\chi(x)=\chi(y)$, so the induced base point is $\chi(y)\in\chi(V)$. Hence
\begin{align*}
B_U(\mathcal{O}_1,\mathcal{O}_2)=\chi(V)=U_0.
\end{align*}
  Thus, we take $\Phi_B:U_0\to B_U(\mathcal{O}_1,\mathcal{O}_2)$ to be this canonical identification. Since the coadjoint quotient carries the trivial Poisson structure, $\Phi_B$ is Poisson.

Finally, we verify the commutativity of the diagram separately as follows:

For the left square, note that $\Phi_S^{-1}:(T^*M,\omega_\varepsilon)\vert_U\to  S_U(\mathcal{O}_1,\mathcal{O}_2)$ and $\Pi_1:U\to U_1$, and the commutativity identity is $\Phi_P\circ \Pi_1 = p_1\circ \Phi_S^{-1}$. 

For the right square, let $(x,[y]_{A})\in U_1$ be a point of the fiber product. Then $\chi(x)=\chi(y)$, and therefore
\begin{align*}
(p_2 \circ \Phi_P)(x,[y]_{A}) = p_2([(x,y)]) = \chi(x) = \chi(y) = (\Phi_B \circ \Pi_2)(x,[y]_{A}).
\end{align*}
Hence, $\Phi_B \circ \Pi_2 = p_2 \circ \Phi_P$. Thus, both squares commute, and the theorem follows.
\end{proof}

%\subsubsection{A special case} With $(K_1,K_2)=({e},A)$ and $(\mathcal{O}_1,\mathcal{O}_2)=({0},\{\varepsilon \gamma\})$: \begin{itemize}
%\item The twist parameter $\varepsilon \gamma$ enters only through the \textit{right} orbit pairs $\mathcal{O}_2=\{\varepsilon \gamma\}\subset\mathfrak{a}^*$, which is a singleton because $A$ centralizes $\gamma$.
%\item The affine slice $\mathfrak{m}-\varepsilon \gamma$ appears because the right constraint $\mu_A^R=\varepsilon \gamma$ is equivalent to $\mathrm{Ad}_{g^{-1}}^*\xi\in\mathfrak{m}-\varepsilon \gamma$.
%\item The intermediate fiber product is the space of \textit{compatible pairs} $(x,[y]_A)$ with $x\in\mathrm{Ad}(G)(\mathfrak{m}-\varepsilon \gamma)$, $y\in\mathfrak{m}-\varepsilon \gamma$, and $\chi(x)=\chi(y)$ (or equivalently, $\mathcal{O}_G(x)=\mathcal{O}_G(y)$ on the regular locus).
%\end{itemize}

%If we use the alternative convention for the right moment map $\mu^R_K(g, \xi)= + \pi_{\mathfrak{k}}(\mathrm{Ad}_{g^{-1}}^*\xi)$ (or equivalently swaps $\{\cdot,\cdot\}_ + $ and $\{\cdot,\cdot\}_-$ in the target), then the correct orbit pairs becomes $\mathcal{O}_2=\{-\varepsilon \gamma\}$. 
\subsection{Coisotropic reduction in $T^*(G/T)$ via the Poisson projections}
\label{subsec:coisotropy}

In this subsection, we are in the regular centralizer case $A=T=Z_G(\gamma)$. That is, $\gamma$ is regular. We start with the magnetic phase space $T^*(G/T)$ and perform its additional left $T$-reduction. To distinguish the left and right factors in the left-right action on $T^*G$, write $T_L:=T\times\{e\}$ and $T_R:=\{e\}\times T$. Note that the action is not free, (the stabilizer of $e$ is $\Delta T$. %That is, the stabilizer of $e$ is the diagonal $\Delta T$.
Hence, Proposition \ref{prop:superintchainso1o2}, which assumes a free and proper action, cannot be applied directly. We apply it only on the subset $\widetilde S$ defined in \eqref{eq:Stilde} below, on which $(T_L\times T_R)/\Delta Z(G)$ acts freely and properly.

The identification with $S_{(T,T)}(0,\varepsilon\pi_{\mathfrak{t}}(\lambda))$ provided below is a reduction by stages argument \cite{marsden2007hamiltonian}: the first stage reduces $T^*G$ by the right factor $T_R$ at level $\varepsilon\pi_{\mathfrak{t}}(\lambda)$ to obtain the magnetic cotangent model $(T^*(G/T),\omega_\varepsilon)$, and the second stage reduces by the residual effective left action of $T_L/Z(G)=T_{\mathrm{eff}}$ on $L$. In particular, $\gamma\in\mathfrak{t}$, $\lambda=B^\flat( \gamma)$, and $\mathfrak{g}=\mathfrak{t}\oplus\mathfrak{m}$. Recall that  $T^*(G/T)\cong G\times_T\widetilde{\mathfrak{t}}$, where $\widetilde{\mathfrak{t}}=\ker\pi_{\mathfrak{t}}$ is the annihilator of $\mathfrak{t}$ in $\mathfrak{g}^*$ (and in this regular case $\widetilde{\mathfrak{t}}=\widetilde{\mathfrak{a}}$). We keep the variable $X\in\widetilde{\mathfrak{t}}$ and write points as equivalence classes $[g,X]_T$ with \begin{align*}
 [g,X]_T=[gt,\mathrm{Ad}_{t^{-1}}^* X]_T,\qquad t\in T.
\end{align*} Since $gt_0T=gT$ for all $t_0\in T$, right multiplication by $T$ on the base $G/T$ is trivial, so its cotangent lift is the identity:
\begin{align}
    [g,X]_T\cdot t_0=[g,X]_T.
\label{eq:rightaction}
\end{align}
Conversely, the left action $t \cdot (gT) = tgT$ is trivial on $G/T$ if and only if $g^{-1}tg \in T$ for all $g \in G$. Since the conjugate maximal tori $gTg^{-1}$ intersect precisely in the center $Z(G)$, the ineffectiveness kernel of this action is $Z(G)$. Moreover, for any $z \in Z(G)$ and any class $[g, X]_T \in T^*(G/T)$,
\begin{align*}
z \cdot [g, X]_T = [zg, X]_T = [gz, X]_T = [g, \mathrm{Ad}_z^* X]_T = [g, X]_T,
\end{align*}
demonstrating that $Z(G)$ acts trivially on all of $T^*(G/T)$. Since $G$ is compact and connected, $Z(G)\subset T$. It is finite since $G$ is semisimple. We define the effective torus
\begin{align*}
T_{\mathrm{eff}} := T / Z(G).
\end{align*}
On $L$ defined in \eqref{eq:L-def}, the stabilizer in $T$ of $[g, X]_T$ is $Z(G)$, so the induced $T_{\mathrm{eff}}$-action is free, as verified in Proposition \ref{prop:thetaiso}.

For any $t_0 \in T$, we define the Hamiltonian left action as follows:
\begin{align}
 t_0 \cdot [g, X]_T := [t_0 g, X]_T. \label{eq:leftaction}
\end{align}
In contrast, on the unreduced cotangent bundle $T^*G$, the diagonal subgroup $\Delta Z(G) := \{(z, z) : z \in Z(G)\}$ forms the ineffective kernel of the joint left-right action, while the individual left and right translations are free.

\begin{definition}\label{def:momentmap}
The moment map for the Hamiltonian left $G$-action on $(T^*(G/T), \omega_\varepsilon)$ is given by
\begin{align*}
J_G([g, X]_T) = \mathrm{Ad}_g^*(X - \varepsilon \lambda) \in \mathfrak{g}^*.
\end{align*}
Restricting this action to the maximal torus $T$ yields the moment map for the left $T$-action on $T^*(G/T)$:
\begin{align}
I([g, X]_T) := (\pi_{\mathfrak{t}} \circ J_G)([g, X]_T) = \pi_{\mathfrak{t}}\bigl(\mathrm{Ad}_g^*(X - \varepsilon \lambda)\bigr) \in \mathfrak{t}^*, \label{eq:leftmoment}
\end{align}
which satisfies $\iota_{\widehat{\eta}} \omega_\varepsilon = d \langle I, \eta \rangle$ for all $\eta \in \mathfrak{t}$, where $\pi_{\mathfrak{t}} : \mathfrak{g}^* \to \mathfrak{t}^*$ is the canonical dual projection.
\end{definition}

\begin{remark}
In these coordinates, we write $J_G(g,X)$ for $J_G([g,X]_T)$.
\end{remark}

We consider the level set $I^{-1}(0)$. Define
\begin{align}
L := \bigl\{[g, X]_T \in T^*(G/T) : I([g, X]_T) = 0, \ X - \varepsilon \lambda \in V, \ T_{J_G(g,X)} = T_{X - \varepsilon \lambda} = Z(G)\bigr\}. \label{eq:L-def}
\end{align}
Since $X-\varepsilon\lambda\in V$ already implies $T_{X-\varepsilon\lambda}=Z(G)$, we can write
\begin{align*}
L=\bigl\{[g,X]_T\in U:\ I([g,X]_T)=0,\; T_{J_G(g,X)}=Z(G)\bigr\}.
\end{align*}
Assume $L\neq\varnothing$. % On the open subset $L \subset I^{-1}(0)$ defined below,
The differential $dI$ has full rank, and the action of $T_{\mathrm{eff}}$ is free.  Consequently, $L$ is a coisotropic submanifold with a characteristic distribution \begin{align*}
(T_z L)^{\omega_\varepsilon} = \{\widehat{\eta}(z) : \eta \in \mathfrak{t}\}, \qquad z \in L,
\end{align*}
which is tangent to the $T$-orbits. By Marsden-Weinstein reduction \cite{MR0402819}, the quotient $M_0 := L/T \cong L/T_{\mathrm{eff}}$ inherits a smooth symplectic structure.

For the factors $T_L,T_R$ fixed above, the moment maps in the right trivialization \eqref{eq:right-trivialization} are
\begin{align*}
\mu_T^L(g, \xi) = \pi_{\mathfrak{t}}(\xi), \qquad \mu_T^R(g, \xi) = -\pi_{\mathfrak{t}}(\mathrm{Ad}_{g^{-1}}^* \xi),
\end{align*}
with joint moment map $\mu_{12} := (\mu_T^L, \mu_T^R) : T^*G \to \mathfrak{t}^* \oplus \mathfrak{t}^*$. For the singleton coadjoint orbits $\{0\}$ and $\{\varepsilon \pi_{\mathfrak{t}}(\lambda)\}$, we define the biquotient
\begin{align}
S_{(T,T)}(0, \varepsilon \pi_{\mathfrak{t}}(\lambda)) &:= T_L \backslash \bigl\{(g, \xi) \in T^*G : \mu_T^L(g, \xi) = 0, \ \mu_T^R(g, \xi) = \varepsilon \pi_{\mathfrak{t}}(\lambda)\bigr\} / T_R \nonumber \\
&\phantom{:}= T_L \backslash C(0, \varepsilon \pi_{\mathfrak{t}}(\lambda)) / T_R. \label{eq:biquotient-level}
\end{align}
Let $V \subset \widetilde{\mathfrak{t}} - \varepsilon \lambda$ be the $T$-invariant open regular subset fixed before Theorem \ref{thm:superin}. We denote
\begin{align}
\widetilde{S} := \bigl\{(g, \xi) \in C(0, \varepsilon \pi_{\mathfrak{t}}(\lambda)) : [g, \mathrm{Ad}_{g^{-1}}^* \xi + \varepsilon \lambda]_T \in L\bigr\}, \qquad S_{(T,T)}^L(0, \varepsilon \pi_{\mathfrak{t}}(\lambda)) := T_L \backslash \widetilde{S} / T_R. \label{eq:Stilde}
\end{align}
For any $\zeta \in \mathfrak{g}^*$, let $T_\zeta := \{t \in T : \mathrm{Ad}_t^* \zeta = \zeta\}$. On the principal orbit-type locus $\widetilde{\mathfrak{t}}_{\mathrm{pr}}$ (and of $\widetilde{\mathfrak{t}}_{\mathrm{pr}} - \varepsilon \lambda$), the stabilizer is $Z(G)$, since a generic element has non-vanishing projections onto every root space.

\begin{proposition}\label{prop:thetaiso}
Let $V \subset \widetilde{\mathfrak{t}} - \varepsilon \lambda$ be the $T$-invariant open subset defined above.  Then:
\begin{enumerate}
    \item[(i)] $L \subset (T^*(G/T), \omega_\varepsilon)$ is a smooth coisotropic submanifold of codimension $\dim T_{\mathrm{eff}}$, and the effective left action of $T_{\mathrm{eff}} = T/Z(G)$ on $L$ is free and proper.
    \item[(ii)] The Marsden-Weinstein reduced space $M_0 := L/T$ is a smooth symplectic manifold of dimension $2\dim G - 4\mathrm{rank} G$.
    \item[(iii)] There is a canonical symplectomorphism
    \begin{align}
    \Theta : M_0 \xrightarrow{\sim} S_{(T,T)}^L(0, \varepsilon \pi_{\mathfrak{t}}(\lambda)), \label{eq:thetamap}
    \end{align}
    given explicitly by
    \begin{align}
    \Theta(T \cdot [g, X]_T) = T_L \cdot \bigl(g, \mathrm{Ad}_g^*(X - \varepsilon \lambda)\bigr) \cdot T_R, \label{eq:theta-explicit}
    \end{align}
 with a smooth inverse
    \begin{align}
    \Theta^{-1}(T_L \cdot (g, \xi) \cdot T_R) = T \cdot [g, \mathrm{Ad}_{g^{-1}}^* \xi + \varepsilon \lambda]_T. \label{eq:thetainverse}
    \end{align}
\end{enumerate}
\end{proposition}

\begin{remark}
Throughout this subsection, we keep the same choice of $V$ made before Theorem \ref{thm:superin} in the case $A=T$. We work with the open subset $L\subset I^{-1}(0)$ defined in \eqref{eq:L-def} and the corresponding open subset $\widetilde{S}\subset C(0,\varepsilon\pi_{\mathfrak{t}}(\lambda))$ defined in \eqref{eq:Stilde}. Their quotients are $M_0=L/T$ and $S^L_{(T,T)}(0,\varepsilon\pi_{\mathfrak{t}}(\lambda))=T_L\backslash\widetilde{S}/T_R$, and the intermediate and base spaces $P_{(T,T)}$ and $B_{(T,T)}$ are defined below using this same $V$.
\end{remark}

\begin{proof}
We first justify items (i)-(ii) on $L$. Since $\lambda$ is $T$-fixed, we have \begin{align*}
T_X=T_{X-\varepsilon\lambda}=Z(G)
\end{align*}
for every $[g,X]_T\in L$. Therefore, the stabilizer of $[g,X]_T$ for the $T$-action is \begin{align*}
\mathrm{Stab}_T([g,X]_T)=T\cap gT_Xg^{-1}=Z(G).
\end{align*} Since $Z(G)$ is finite, the moment map identity $\iota_{\widehat\eta}\omega_\varepsilon=d\langle I,\eta\rangle$ implies that \begin{align*}
\mathrm{rank}(dI)_z=\dim T\qquad\text{for every }z\in L.
\end{align*}
Thus, $L\subset I^{-1}(0)$ is open and consists of regular points of $I$, and its characteristic distribution is spanned by the infinitesimal $T$-action directions. The induced action of $T_{\mathrm{eff}}=T/Z(G)$ on $L$ is therefore free, and the properness is automatic due to the compactness. Finally,
\begin{align*}
\dim(L/T)=2(\dim G-\dim T)-2\dim T=2\dim G-4\mathrm{rank}G.
\end{align*}

We now prove item (iii). By Proposition \ref{prop:mincoup} (with $A=T$), $T^*(G/T)$ is symplectomorphic to the reduction
\begin{align*}
    C_\varepsilon/T_R,\qquad C_\varepsilon:=(\mu_T^R)^{-1}(\varepsilon\pi_{\mathfrak{t}}(\lambda))\subset T^*G.
\end{align*}
Define
\begin{align*}
\Psi_{R}:T^*(G/T)\to C_\varepsilon/T_R,
\qquad \Psi_{R}([g,X]_T):=[g,\mathrm{Ad}_g^*(X-\varepsilon\lambda)]_{T_R},
\end{align*}
and
\begin{align*}
\Phi_R:C_\varepsilon/T_R\to T^*(G/T),
\qquad \Phi_{R}([g,\xi]_{T_R}):=[g,\mathrm{Ad}_{g^{-1}}^*\xi+\varepsilon\lambda]_T.
\end{align*}
By Proposition \ref{prop:mincoup} and Remark \ref{rem:homogeneouscoord}, these maps are mutually inverse symplectomorphisms. Under $\Psi_R$, the left $T$-action \eqref{eq:leftaction} corresponds to the induced left action on $C_\varepsilon/T_R$, whose moment map is the restriction of the standard left moment map.
As the left and right $T$-actions on $T^*G$ commute, the left action descends to $C_\varepsilon/T_R$. Explicitly,  $a\cdot [g,\xi]_{T_R}:=[ag,\mathrm{Ad}_{a}^*\xi]_{T_R}$.
Its moment map is \begin{align*}
   \widetilde{I}: C_\varepsilon/T_R\longrightarrow \mathfrak{t}^*, \qquad \widetilde{I}([g,\xi]_{T_R}):=\mu_T^L(g,\xi)= \pi_{ \mathfrak{t}} (\xi),
\end{align*}
where $\pi_{\mathfrak{t}}:\mathfrak{g}^*\to\mathfrak{t}^*$ is restriction. We now compare $\widetilde I$ with the moment map $I$ on $T^*(G/T)$.
For every $[g,X]_T\in T^*(G/T)$, \begin{align*} 
\nonumber
    (\widetilde{I}\circ \Psi_{R})([g,X]_T) = & \widetilde{I}\bigl([g,\mathrm{Ad}_g^* (X-\varepsilon \lambda )]_{T_R}\bigr)  \\
    = &   \pi_{ \mathfrak{t}} \bigl(\mathrm{Ad}_g^* (X-\varepsilon \lambda )\bigr) = I([g,X]_T).
\end{align*} Hence, $\widetilde{I}\circ \Psi_{R}=I$. 

Moreover, $\Psi_{R}$ intertwines the left $T$-actions.
Indeed, \begin{align*}
    \Psi_{R}\bigl(a \cdot [g,X]_T\bigr)   %& \Psi_{R}\bigl([ag,X]_T\bigr) \\
  %  = & [ag,\mathrm{Ad}_{ag}(X-\varepsilon \gamma)]_{T_R} \\
    =   a  \cdot[g,\mathrm{Ad}_g^*(X-\varepsilon\lambda)]_{T_R}%\\ 
    =   a\cdot\Psi_{R}([g,X]_T).
\end{align*} Therefore, $\Psi_{R}$ restricts to a $T$-equivariant diffeomorphism  $L\xrightarrow{\ \sim\ }\Psi_R(L)$, and hence induces a diffeomorphism   $L/T \xrightarrow{\ \sim\ } \Psi_R(L)/T_{L}$.
By the definitions of $\widetilde{S}$ and $\Psi_R$,\begin{align*}
\Psi_R(L) = \bigl\{[g,\xi]_{T_R}:(g,\xi)\in \widetilde{S}\bigr\} = \widetilde{S}/T_R.
\end{align*}  Define \begin{equation}
  \Xi: \Psi_R(L)/T_{L}\longrightarrow T_{L}\backslash \widetilde{S}/T_R, \qquad \Xi\bigl(T_{L}\cdot[z]_{T_R}\bigr):=T_{L}\cdot z\cdot T_R.
\label{eq:inversexi}
\end{equation} 
The map $\Theta$ in \eqref{eq:thetamap} is the composition of the map induced by $\Psi_R$ with $\Xi$.
We check that $\Xi$ is well defined. If $[z_1]_{T_R}=[z_2]_{T_R}$, then $z_2=z_1\cdot t$ for some $t\in T_R$, hence \begin{align*}
  T_{L}\cdot z_2\cdot T_R=T_{L}\cdot z_1\cdot T_R.
\end{align*} %If, in addition, $[z_2]_{T_R}=a\cdot[z_1]_{T_R}$ for some $a\in T_{L}$, then \begin{align*}
 % T_{L}\cdot z_2\cdot T_R=T_{L}\cdot z_1\cdot T_R.
%\end{align*} 
So $\Xi$ does not depend on any choices of $[z]_{T_R}$. Now, we need to show that $\Xi$ is bijective.
We first show that it is surjective. Every double class $T_{L}\cdot z\cdot T_R$ with $z\in \widetilde{S}$ is the image of $T_{L}\cdot[z]_{T_R}$.
On the other hand, it is injective. Indeed, if \begin{align*}
    \Xi\bigl(T_{L}\cdot[z_1]_{T_R}\bigr) = \Xi\bigl(T_{L}\cdot[z_2]_{T_R}\bigr),
\end{align*} then $T_{L}\cdot z_1\cdot T_R=T_{L}\cdot z_2\cdot T_R$.
Hence, there exist $a\in T_{L}$ and $t\in T_R$ such that $z_2=a\cdot z_1\cdot t$.
Proceeding to the right quotient gives $ [z_2]_{T_R}=a\cdot[z_1]_{T_R}$, hence  $T_{L}\cdot[z_1]_{T_R}=T_{L}\cdot[z_2]_{T_R}$.
Therefore, $\Xi$ is a bijection, and it is a diffeomorphism because the quotient maps for the free and proper effective actions are surjective submersions, and both $\Xi$ and its inverse are induced by the identity on $\widetilde S$. Hence, we obtain \begin{align*}
L/T\cong\Psi_R(L)/T_L\cong T_L\backslash \widetilde{S}/T_R =S_{(T,T)}^L(0,\varepsilon\pi_{\mathfrak{t}}(\lambda)).
\end{align*} 
This proves the claimed identification.

Finally, we show the symplectic form is preserved.
Since the left and right translation actions of $T_L$ and $T_R$ on $T^*G$ commute, the individual actions are free, and the joint action has ineffective kernel $\Delta Z(G)$. On the selected level subset $\widetilde{S}\subset (\mu_T^L)^{-1}(0)\cap(\mu_T^R)^{-1}(\varepsilon\pi_{\mathfrak{t}}(\lambda))$, the corresponding joint effective group $(T_L\times T_R)/\Delta Z(G)$ acts freely and properly.

Reduction by stages therefore takes the form: first reduce by $T_R$ at level $\varepsilon\pi_{\mathfrak{t}}(\lambda)$ to obtain $S:=C_\varepsilon/T_R$, and then reduce by the residual effective group $T_L/Z(G)=T_{\mathrm{eff}}$ at level $0$. In particular, the reduced symplectic forms agree with those obtained by reducing by the joint effective group.

Denote by \begin{align*}
    \pi_R:\widetilde{S}\to \widetilde{S}/T_R,\qquad \pi_L:\widetilde{S}/T_R\to T_L\backslash \widetilde{S}/T_R
\end{align*} the quotient maps (equivalently, the second stage is the quotient by $T_{\mathrm{eff}}$), and by $i:\widetilde{S}\hookrightarrow T^*G$ the inclusion. The first-stage symplectic manifold is $S:=C_\varepsilon/T_R$, containing $\widetilde{S}/T_R$ as the residual moment-map level subset. Let $\omega_S$ be its reduced symplectic form and $j:\widetilde{S}/T_R=\Psi_R(L)\hookrightarrow S$ the inclusion. The form $\omega_{\widetilde{S}/T_R}:=j^*\omega_S$ is presymplectic and satisfies
\begin{align*}
\pi_R^*\omega_{\widetilde{S}/T_R}=i^*\omega_{G,\mathrm{can}},\qquad \pi_L^*\omega_{\mathrm{red}}=\omega_{\widetilde{S}/T_R}.
\end{align*}
The second identity is Marsden-Weinstein reduction of the Hamiltonian left $T_{\mathrm{eff}}$-action on $S$ at zero. The symplectomorphism $\Phi_R:S\to(T^*(G/T),\omega_\varepsilon)$ takes $\widetilde{S}/T_R$ to $L$ and is equivariant for the residual effective group. Let $q_0:L\to L/T$ be the quotient map and let $\omega_0$ be its reduced symplectic form. The surjective submersion \begin{align*}
f:=q_0\circ\Phi_R\circ\pi_R:\widetilde{S}\longrightarrow L/T
\end{align*}
satisfies $\Theta\circ f=\pi_L\circ\pi_R$. Therefore, \begin{align*}
f^*(\Theta^*\omega_{\mathrm{red}})  &=(\pi_L\circ\pi_R)^*\omega_{\mathrm{red}}   =i^*\omega_{G,\mathrm{can}}   = f^* \omega_0.
\end{align*}
the injectivity of $f^*$ gives $\Theta^*\omega_{\mathrm{red}}=\omega_0$.

Consequently, $\Theta$ identifies the reduced symplectic forms and is a symplectomorphism. %Composing the stage maps yields the explicit formulas in \eqref{eq:Thetavalue}.
\end{proof} 

\begin{remark}
When $\varepsilon=0$, the cotangent-bundle identification applies over an invariant open subset of $G$ on which $(T\times T)/\Delta Z(G)$ acts freely. On that subset, the zero-level reduction is $T^*(T\backslash G/T)$. %The smoothness of a cotangent-level quotient alone does not provide this identification over the entire biquotient. 
For $\varepsilon\neq0$, we obtain the corresponding magnetic reduction. %  On the regular locus, we can identify $I^{-1}(0)/T $ by $T^*(T\backslash G/T)$.
In particular,
\begin{align*}
    S_{(T,T)}(0,\varepsilon \pi_{\mathfrak{t}}(\lambda) ) = T_L \big\backslash \left((\mu_T^L)^{-1}(0) \cap (\mu_T^R)^{-1}(\varepsilon \pi_{\mathfrak{t}}(\lambda) )\right) \big/T_R.
\end{align*}
The reduction in stages gives
\begin{align}
    S_{(T,T)}(0,\varepsilon \pi_{\mathfrak{t}}(\lambda) ) \cong \bigl(T^*G//_{\varepsilon \pi_{\mathfrak{t}}(\lambda)} T_R\bigr)//_0 T_L.
\end{align}
\end{remark}

%{\color{blue} Adding a paragraph to explain here we have already use the regular locus to define the intermediate space $\mathfrak{g} \times_{\mathrm{Spec}\,R_0} \mathfrak{m}//T$.
%See, \cite{jiang2026poisson} for the reason why we ignore the regular notation.

Now, define the quotient maps \begin{align}
    \kappa_0:    \widetilde{\mathfrak{t}} \longrightarrow \widetilde{\mathfrak{t}}  //T, \qquad  \kappa_\varepsilon:   \widetilde{\mathfrak{t}}  - \varepsilon  \lambda \longrightarrow (\widetilde{\mathfrak{t}} -\varepsilon \lambda ) //T.
\label{eq:kappa}
\end{align} As $T$ is compact, its $T$-orbits are closed, and the equality of the quotient classes is equivalent to lying in the same $T$-orbit.
Note that the translation by $-\varepsilon\lambda$ is $T$-equivariant \cite{jiang2026poisson}. Therefore,  $(\widetilde{\mathfrak{t}} -\varepsilon\lambda) //T$ can be identified with $\widetilde{\mathfrak{t}}//T$. Here $\widetilde{\mathfrak{t}}_{\mathrm{pr}}$ denotes the principal orbit-type locus for the coadjoint $T$-action on $\widetilde{\mathfrak{t}}$. The open subset $V$ remains fixed.

Define \begin{align*}
P_{(T,T)}:=\Bigl\{(\kappa_0(Y),\kappa_\varepsilon(Z)):\;  Y\in\widetilde{\mathfrak{t}} ,\quad Z\in V, \chi(Y)=\chi(Z) ,\quad T_Y=T_Z=Z(G)\Bigr\}.
\end{align*}
Moreover, \begin{align*}
P_{(T,T)}\cong\kappa_0(\widetilde{\mathfrak{t}}_{\mathrm{pr}}\cap\mathfrak{g}^*_{\mathrm{reg}})\times_{\chi(\mathfrak{g}^*_{\mathrm{reg}})}\kappa_\varepsilon(V).
\end{align*}
This is precisely the intermediate manifold $P(\mathcal{O}_1,\mathcal{O}_2)$ from \eqref{eq:intermide} in the special case $(K_1,K_2)=(T,T)$ and $(\mathcal{O}_1,\mathcal{O}_2)=(\{0\},\{\varepsilon\pi_{\mathfrak{t}}(\lambda)\})$, restricted to $\kappa_0\bigl(\widetilde{\mathfrak{t}}_{\mathrm{pr}}\cap\mathfrak{g}^*_{\mathrm{reg}}\bigr)$ and $\kappa_{\varepsilon}(V)$ (with $T_Y=T_Z=Z(G)$).
Similarly, define \begin{align*}
B_{(T,T)}:=\chi\Bigl(\Bigl\{Z\in V:\;&
\text{there exists }Y\in\widetilde{\mathfrak{t}} \cap\mathfrak{g}^*_{\mathrm{reg}}\\
&\text{with }\chi(Y)=\chi(Z),\ T_Y=T_Z=Z(G)\Bigr\}\Bigr).
\end{align*} Similar to the case above, we observe that \begin{align*}
B_{(T,T)}
&=\chi\Bigl(\Bigl\{Z\in V:\ T_Z=Z(G), \ Z\in\mathrm{Ad}^* (G)\{Y\in\widetilde{\mathfrak{t}} :T_Y=Z(G)\}\Bigr\}\Bigr),
\end{align*} which is the specialized case in \eqref{eq:base}. %(\rev{restricted to the subsets just defined}).
%The map $q_P^R$ is immediate from the definition of the quotient maps \begin{align*} \kappa_0: \mathfrak{m} \to \mathfrak{m} /T, \qquad \kappa_{\varepsilon}: \mathfrak{m}-\varepsilon \gamma \to \mathfrak{m}-\varepsilon \gamma /T.\end{align*} 

On these free coordinate sets, the rank computation preceding Theorem \ref{thm:superin} gives a rank of $r:=\mathrm{rank}G$ for both restrictions of $\chi$. Their images are open in $\chi(\mathfrak{g}^*_{\mathrm{reg}})$, so $B_{(T,T)}$ is their open intersection. The free quotients and their fiber product are therefore smooth. Writing $d=\dim G$, their dimensions are \begin{align*}
\dim M_0=2d-4r,\qquad \dim P_{(T,T)}=2d-5r,\qquad \dim B_{(T,T)}=r.
\end{align*}
Indeed, each affine coordinate set has dimension $d-r$, its free torus quotient has dimension $d-2r$, and the fiber product subtracts $r$. The phase-space dimension follows from the regular effective left $T$-reduction of $T^*(G/T)$. Thus, the dimension identity required in Definition \ref{def:superchain1} holds on this locus.

In the following theorem, we aim to build the superintegrable Poisson projection chain on the phase space $M_0  = L /T$.

\begin{lemma}
\label{lem:TT-leaves}
Let $b\in B_{(T,T)}$. Then every connected component of $(\pi_2^T)^{-1}(b)$ is a symplectic leaf of $P_{(T,T)}$.
\end{lemma}

\begin{proof}
Fix $b\in B_{(T,T)}$.
Since $b\in B_{(T,T)}\subset\chi(\mathfrak{g}^*_{\mathrm{reg}})$, the fiber of the coadjoint quotient over $b$ is a single regular coadjoint orbit $\mathcal{O}_b$:
\begin{align*}
\chi^{-1}(b)\cap \mathfrak{g}^* _{\mathrm{reg}}  = \mathcal{O}_b,
\end{align*}
where $\chi:  \mathfrak{g}^* \rightarrow \mathfrak{g}^*//G$.
Hence, the fiber of $\pi_2^T$ over $b$ may be identified with
\begin{align}
\bigl(\pi_2^T\bigr)^{-1}(b) \cong \left\{(Y,Z)\in\mathcal{O}_b\times\mathcal{O}_b^-:
\begin{array}{l}
\pi_{\mathfrak t}(Y)=0,\quad Z\in V,\\
T_Y=T_Z=Z(G)
\end{array}\right\}\Big/(T\times T). \label{eq:TT-fiber-reduction}
\end{align}
This is the part of the reduction of $\mathcal{O}_b\times\mathcal{O}_b^-$ at $(0,\varepsilon\pi_{\mathfrak t}(\lambda))$ represented by $Z\in V$ and $T_Y=T_Z=Z(G)$. Here, $\mathcal{O}_b^-$ carries the opposite KKS form.

The product $\mathcal{O}_b\times \mathcal{O}_b^{-}$ is symplectic, and the $T\times T$-action is Hamiltonian with moment map \begin{align*}
\mu_b: \mathcal{O}_b\times \mathcal{O}_b^{-}\longrightarrow \mathfrak{t}^*\times \mathfrak{t}^*, \qquad \mu_b(Y,Z):= \bigl(\pi_{\mathfrak{t}}(Y),-\pi_{\mathfrak{t}}(Z)\bigr).
\end{align*}
By the regularity and effective freeness on the selected portion, the reduced space in \eqref{eq:TT-fiber-reduction} is smooth, and each connected component carries the reduced symplectic form. Consequently, each connected component of $\bigl(\pi_2^T\bigr)^{-1}(b)$ is a connected symplectic manifold.

On the other hand, $P_{(T,T)}$ carries the quotient Poisson structure, and the map $\pi_2^T$ is Poisson by construction. Locally, a pullback of a function on $B_{(T,T)}$ is induced by a smooth coadjoint-invariant function of $Y$, hence by a Casimir on $\mathfrak{g}^*$. It remains central after reduction. Consequently, all Hamiltonian vector fields on $P_{(T,T)}$ are tangent to the fibers of $\pi_2^T$.

Thus, each fiber of $\pi_2^T$ is a union of symplectic leaves. Since each connected component $C$ of the fiber is already symplectic, the induced Poisson tensor on $C$ is nondegenerate and hence agrees with the reduced symplectic form. Hence, every leaf meeting a component $C$ of the fiber is open in $C$. Since $C$ is connected, $C$ is a single leaf.
\end{proof}

\begin{theorem}
\label{thm:superinchaini}
Let $L$, $M_0=L/T$, and $\Theta$ be as in Proposition \ref{prop:thetaiso}. Then \begin{align}
M_0 \xrightarrow{\; \pi_1^T \; } P_{(T,T)} \xrightarrow{\;\pi_2^T\; } B_{(T,T)} \label{eq:poissioni10}
\end{align} is a superintegrable  Poisson projections chain, where $\pi_1^T=p_1\circ\Theta$ and $\pi_2^T=p_2$, and where the canonical projections are \begin{align}
      p_1([g,\xi])=[(\xi,\mathrm{Ad}_{g^{-1}}^*\xi)],\qquad p_2([(Y,Z)])=\chi(Y)=\chi(Z).
\end{align}
\end{theorem}

\begin{remark}
    In terms of representatives, if a point of $M_0$ is represented by $[g,X]_T\in L$, then \begin{align*}
 \pi_1^T\bigl(T\cdot [g,X]_T\bigr)  =\bigl(\kappa_0(J_G(g,X)),\kappa_{\varepsilon}(X-\varepsilon \lambda )\bigr),\qquad  \pi_2^T\bigl(\kappa_0(Y),\kappa_{\varepsilon}(Z)\bigr)  = \chi(Y),
\end{align*}
where $\kappa_0$ and $\kappa_\varepsilon$ are defined in \eqref{eq:kappa}.
\end{remark}

\begin{proof}
We apply the equivariance and Poisson-reduction argument of Proposition \ref{prop:superintchainso1o2} to $\widetilde S\subset C(0,\varepsilon\pi_{\mathfrak{t}}(\lambda))$, with quotient $S^L_{(T,T)}(0,\varepsilon\pi_{\mathfrak{t}}(\lambda))$ and target spaces $P_{(T,T)}$ and $B_{(T,T)}$ defined above.

For $\mathcal{O}_1=\{0\}$ and $\mathcal{O}_2=\{\varepsilon\pi_{\mathfrak t}(\lambda)\}$, the induced Poisson maps are \begin{align*}
S_{(T,T)}^L (0,\varepsilon \pi_{\mathfrak{t}}(\lambda) ) \xrightarrow{p_1} P_{(T,T)} \xrightarrow{p_2} B_{(T,T)}.
\end{align*}
Since $\Theta$ is a symplectomorphism, $\pi_1^T=p_1\circ\Theta$ and $\pi_2^T=p_2$ are Poisson. We verify their surjectivity and submersion below on these domains. The symplectic leaf condition is established in Lemma \ref{lem:TT-leaves}.

Moreover, any pullback $\bigl(\pi_2^T\bigr)^*f$ with $f\in C^\infty(B_{(T,T)})$ is induced by a coadjoint-invariant function and hence is a Casimir after reduction. Therefore, for all $f\in C^\infty(B_{(T,T)})$ and $g\in C^\infty(P_{(T,T)})$, \begin{align*}
\{\bigl(\pi_2^T\bigr)^*f,\,g\}_{P_{(T,T)}}=0.
\end{align*}
Since $\pi_1^T$ is Poisson, pulling back to $M_0$ yields \begin{align*}
\{\bigl(\pi_2^T\circ\pi_1^T\bigr)^*f,\,\bigl(\pi_1^T\bigr)^*g\}_{M_0}=0.
\end{align*}
Thus, the abelian algebra of commuting Hamiltonians $\bigl(\pi_2^T\circ\pi_1^T\bigr)^*C^\infty(B_{(T,T)})$ Poisson-commutes with the full algebra of conserved quantities $\bigl(\pi_1^T\bigr)^*C^\infty(P_{(T,T)})$.

For reference, letting $d:=\dim G$ and $r:=\mathrm{rank}G=\dim T$, we have
\begin{align*}
\dim M_0=2d-4r,\qquad \dim P_{(T,T)}=2d-5r,\qquad \dim B_{(T,T)}=r,
\end{align*}
and hence the Mishchenko-Fomenko dimension criterion \cite{MishchenkoFomenkoGeneralized1978,nehorovsev1968action} holds:
\begin{align*}
\dim P_{(T,T)}+\dim B_{(T,T)}=(2d-5r)+r=2d-4r=\dim M_0.
\end{align*}

The well-definedness of $\pi_1^T$ is immediate: $\Theta$ is a well-defined symplectomorphism by Proposition \ref{prop:thetaiso}, and $p_1$ is well-defined on the reduction by Proposition \ref{prop:superintchainso1o2}. %Hence, their composition $\pi_1^T=p_1\circ\Theta$ is well-defined. 
The displayed formula follows by substituting the explicit definitions of $\Theta$ and $p_1$. Furthermore, the well-definedness of $\pi_2^T$ is immediate from the defining relation $\chi(Y)=\chi(Z)$ in $P_{(T,T)}$.

Finally, $\pi_1^T:M_0\to P_{(T,T)}$ is surjective. Let $\bigl(\kappa_0(Y),\kappa_\varepsilon(Z)\bigr)\in P_{(T,T)}$ with $Y\in\widetilde{\mathfrak{t}}$, $Z\in V$, and $\chi(Y) = \chi(Z)$.
On the regular locus, the fiber of $\chi$ is a single coadjoint orbit, so there exists $g\in G$ such that  $Y=\mathrm{Ad}_g^* Z$. Writing $Z=X - \varepsilon\lambda$, we have $J_G(g,X)=\mathrm{Ad}_g^*(X-\varepsilon\lambda)=Y$, and hence \begin{align*}
I([g,X]_T)= \pi_{ \mathfrak{t}} (J_G(g,X))= \pi_{ \mathfrak{t}} (Y)=0.
\end{align*}
Since $Z\in V$, $J_G([g,X]_T)=Y$, and $T_Y=T_Z=Z(G)$, all conditions in \eqref{eq:L-def} hold. Hence, $[g,X]_T\in L$, and $\pi_1^T\bigl(T\cdot[g,X]_T\bigr)=\bigl(\kappa_0(Y),\kappa_{\varepsilon}(Z)\bigr)$.

We next check that $\pi_1^T$ is a submersion. Before taking quotients, consider the incidence submanifold
\begin{align*}
  \mathcal{I}:=\{(Y,Z)\in \widetilde{\mathfrak{t}}\times V:\ \chi(Y)=\chi(Z),\ T_Y=T_Z=Z(G)\}
\end{align*}
Let $(\dot Y,\dot Z)\in T_{(Y,Z)}\mathcal{I}$, i.e. $d\chi_Y(\dot Y)=d\chi_Z(\dot Z)$. For $Y=\mathrm{Ad}_g^*Z$, the difference $\dot Y-\mathrm{Ad}_g^*\dot Z$ lies in $\ker d\chi_Y=\{\mathrm{ad}_u^*Y:u\in\mathfrak{g}\}$, so it can be produced by varying $g$. Since the residual effective actions are free and proper on this locus, the quotient spaces are smooth manifolds, and the induced map $\pi_1^T:M_0\to P_{(T,T)}$ is a submersion.

The second projection is a submersion since both coordinate quotient maps to the open base are submersions. Together with the dimension calculation and the reduced leaf description, this verifies all the chain conditions. Therefore, on the open locus $L\subset I^{-1}(0)$ (equivalently, on $M_0=L/T$) and the corresponding restricted quotient spaces $P_{(T,T)}$ and $B_{(T,T)}$ determined by $V$ and the stabilizer conditions $T_Y=T_Z=Z(G)$, we obtain the superintegrable Poisson projection chain
\begin{equation*}
M_0 \xrightarrow{\;\pi_1^T\;} P_{(T,T)} \xrightarrow{\;\pi_2^T\;} B_{(T,T)}.
\end{equation*} 
\end{proof}

\section{Conclusion}
\label{sec:conclusion}

In this work, we developed a geometric description of classical superintegrability in terms of Poisson projection chains and the associated comparison notions. Within this framework, we compared two families of superintegrable systems arising from Lie-theoretic constructions: twisted (magnetic) cotangent bundles over homogeneous spaces $G/A$ and left-right Hamiltonian reduction models on $T^*G$ that produce spin Calogero-Moser type systems. Our main results establish, on a common smooth locus, an equivalence between the corresponding chains: the phase spaces are identified by a symplectomorphism, the intermediate Poisson spaces and base (spectral) spaces match, and the Poisson projections commute with these identifications.

Several directions remain open. It would be interesting to extend the equivalence beyond the smooth loci considered here and to incorporate singularities of the adjoint quotient and the biquotient reduction. Another natural problem is to formulate the same comparison at the quantum level, where centralizer algebras and spectral invariants should admit a parallel description.
%\section*{Acknowledgement}

\section*{Acknowledgement}

Kai Jiang was supported by the Fundamental Research Funds for the Central Universities. Guorui Ma was supported by the Research Funding at SIMIS. The authors thank Nicolai Reshetikhin for useful discussions. The authors also thank L{\'a}szl{\'o} Feh\'{e}r, Ian Marquette, Zaijiu Shang and Yao-Zhong Zhang for helpful comments.
We used the Overleaf AI assistant during this revision to help with notation, exposition, and references. %The authors are responsible for reviewing the revisions and for the final mathematical content.

\section*{Conflict of Interest}
The authors declare that there is no conflict of interest.

\section*{Data Availability}
  No data associated in the manuscript.

\appendix

 \section{Closed \texorpdfstring{$2$}{2}-forms from curvature pairing on \texorpdfstring{$Q = K_1 \backslash G/K_2$}{Q = K1\textbackslash G/K2}}
\label{app:twistedform}
We next recall the standard curvature-pairing construction, which produces closed $2$-forms on $Q$ from the principal bundle $p: G\to Q$. In our setting, the free and proper left-right action of $H:=K_1\times K_2$ on $G$ makes $p$ a principal $H$-bundle over the biquotient $Q=K_1\backslash G/K_2$. Here $\mathfrak h:=\mathrm{Lie}(H)$, and the principal right action is $g\cdot(k_1,k_2)=k_1^{-1}gk_2$, i.e. the inverse of the Hamiltonian left action. We begin with the following lemma:

\begin{lemma}
\label{lem:factr}
    Let $p:  V\rightarrow W$ be a linear surjective map and $f \in V^*$ such that $f\vert_{\ker p} = 0$.
Then there exists a unique $\overline{f} \in W^*$ such that $f = \overline{f} \circ p = p^*\overline{f}$.
\end{lemma}

\begin{proof}
Since $p$ is surjective, by definition, for each $w\in W$, there exists a $v\in V$ such that $p(v)=w$ and defines $\overline{f}(w):= f(v)$.
This is well-defined: if $p(v_1)=p(v_2)=w$, then $v_1-v_2\in\ker p$, hence $f(v_1)-f(v_2)=f(v_1-v_2)=0$ by the assumption $f\vert_{\ker p}=0$.
The linearity of $\overline{f}$ follows from the linearity of $f$ and $p$. Moreover, for all $v\in V$, \begin{align*}
    (\overline{f}\circ p)(v) = \overline{f}(p(v)) = f(v),
\end{align*} such that $f=\overline{f}\circ p=p^*\overline{f}$.
For uniqueness, if $g\in W^*$ satisfies $f=g\circ p$, then for any $w\in W$ and any $v\in V$ with $p(v)=w$, we have $g(w)=g(p(v))=f(v)=\overline{f}(w)$.
Hence, $g=\overline{f}$.
\end{proof}

The construction proceeds as follows: one chooses a principal connection $\theta$ on $p$, forms its curvature $F_\theta\in\Omega^2(G,\mathfrak{h})$, and then pairs $F_\theta$ with an $\mathrm{Ad}(H)$-invariant covector $\nu\in(\mathfrak{h}^*)^H$.
This yields an ordinary $2$-form on the total space,
\begin{align*}
\widetilde{\beta}_\nu:=\langle\nu,F_\theta\rangle\in\Omega^2(G),
\end{align*}
which is horizontal and $H$-invariant, hence basic, and therefore descends uniquely to a $2$-form $\beta_\nu$ on $Q$ (by Lemma \ref{lem:factr}, applied pointwise to the surjective map $\Lambda^2(dp)_g:\Lambda^2T_gG\to\Lambda^2T_{p(g)}Q$).
Finally, the Bianchi identity for $F_\theta$ implies that the descended form is closed. For the fixed connection $\theta$, varying $\nu$ produces this family of closed $2$-forms. The forms depend on $\theta$, only their de Rham cohomology classes are independent of the connection. Indeed, for another connection $\theta'$, the basic one-form $\langle\nu,\theta'-\theta\rangle$ descends to a one-form on $Q$, whose exterior derivative is the difference of the two curvature-pairing forms.
\begin{proposition} \label{prop:curvature}
Choose a principal connection $\theta \in \Omega^1(G,\mathfrak{h})$ on the principal $H$-bundle $p: G \to Q$.
Let $F_\theta \in \Omega^2(G, \mathfrak{h})$ denote its curvature. Let $\nu \in (\mathfrak{h}^*)^H$ be an $\mathrm{Ad}(H)$-invariant covector.
Define the ordinary $2$-form on $G$ by  $\widetilde{\beta}_\nu  = \langle \nu, F_\theta\rangle \in \Omega^2(G)$.
Then \begin{enumerate}
\item $\widetilde{\beta}_\nu$ is basic (horizontal and $H$-invariant), hence there exists a unique $2$-form  $\beta_\nu \in \Omega^2(Q)$ such that $p^*\beta_\nu  =   \widetilde{\beta}_\nu$, by Lemma \ref{lem:factr} applied to $\Lambda^2(dp)_g$.
\item The descended form $\beta_\nu$ is closed, i.e., $d\beta_\nu = 0$.
\end{enumerate}
\end{proposition}

\begin{proof}
A principal connection on $p: G\rightarrow Q$ is a $\mathfrak{h}$-valued $1$-form $\theta \in \Omega^1(G, \mathfrak{h})$ satisfying
\begin{align*}
\text{(i)}\ &\theta(\widehat{X})  =  X \quad \text{for all }X \in \mathfrak{h},
\qquad
\text{(ii)}\  (R_h)^*\theta  =  \mathrm{Ad}(h^{-1})\theta \quad \text{for all }h \in  H,
\end{align*}
where $\widehat{X}$ denotes the fundamental vector field generated by $X$ via the right action of $H$ on $G$, and $R_h$ denotes right multiplication by $h$.
The curvature $F_{\theta} \in \Omega^2(G, \mathfrak{h})$ is defined by the structure equation
\begin{align}
F_{\theta} :=  d\theta  +  \frac{1}{2} [\theta\wedge \theta], \label{eq:structureq}
\end{align}
where the bracket wedge is the $\mathfrak{h}$-valued $2$-form defined by
\begin{align*}
[\theta\wedge \theta](u,v):=  [\theta(u),\theta(v)]  -  [\theta(v),\theta(u)] =  2[\theta(u),\theta(v)] \qquad  u,v \in  T_gG,
\end{align*}
so that $\frac{1}{2}[\theta\wedge \theta](u,v)=[\theta(u),\theta(v)]$.

\medskip\noindent
\textbf{Step 1 \& 2: Horizontality and equivariance.} By the standard structure equations of a principal connection,
$$\iota_{\widehat{X}}F_{\theta} = \iota_{\widehat{X}}d\theta + \frac{1}{2}\iota_{\widehat{X}}[\theta\wedge \theta] = -[X,\theta] + [X,\theta] = 0.$$
Furthermore, the equivariance $(R_h)^*\theta = \mathrm{Ad}(h^{-1})\theta$ extends to the curvature, yielding $(R_h)^*F_{\theta} = \mathrm{Ad}(h^{-1})F_{\theta}$.

\medskip\noindent
\textbf{Step 3: $\widetilde{\beta}_\nu$ is basic and hence descends.}
Define  $\widetilde{\beta}_\nu:= \langle \nu, F_{\theta}\rangle \in \Omega^2(G)$.
Then $\widetilde{\beta}_\nu$ is horizontal (it annihilates vertical vectors) and $H$-invariant (since $\mathrm{Ad}(h)^*\nu=\nu$), hence basic.
Therefore, there exists a unique $2$-form $\beta_\nu \in \Omega^2(Q)$ such that $p^*\beta_\nu  =  \widetilde{\beta}_\nu$.
(Here horizontality gives independence of the choice of horizontal lifts, $H$-invariance gives independence of representatives along fibers, and local sections yield smoothness.)

\medskip\noindent
\textbf{Step 4: closedness of $\beta_\nu$.}
We first prove that $\widetilde{\beta}_\nu$ is closed.
Consider the Bianchi identity for the curvature form $F_{\theta}$:
$dF_{\theta}  +  [\theta\wedge F_{\theta}]  =  0$.
Pairing with $\nu$ yields
\begin{align*}
d\widetilde{\beta}_\nu =  d\langle \nu, F_{\theta}\rangle  =  \langle \nu, dF_{\theta}\rangle  =  -\langle \nu, [\theta\wedge F_{\theta}]\rangle.
\end{align*}
We now use the $\mathrm{Ad}(H)$-invariance of $\nu$ to show that $\nu$ vanishes on commutators. Indeed, for any $X,Y\in\mathfrak{h}$, the function
\begin{align*}
 t\longmapsto \nu\bigl(\mathrm{Ad}(\exp(tX))Y\bigr)
\end{align*}
is constant. Differentiating at $t=0$ gives $\nu([X,Y])=0$, and hence $\nu$ annihilates the commutator ideal $[\mathfrak{h},\mathfrak{h}]$.
But pointwise, the $\mathfrak{h}$-valued $3$-form $[\theta\wedge F_{\theta}]$ takes values in the commutator ideal $[\mathfrak{h},\mathfrak{h}]$, since it is built from Lie brackets.
Hence, $\langle \nu, [\theta\wedge F_{\theta}]\rangle = 0$, and therefore $d\widetilde{\beta}_\nu=0$.
Finally, we show $d\beta_\nu=0$ on $Q$. Since $p^*\beta_\nu=\widetilde{\beta}_\nu$, we have \begin{align*}
p^*(d\beta_\nu) =  d(p^*\beta_\nu)  =  d\widetilde{\beta}_\nu  = 0.
\end{align*} Because $p: G\to Q$ is a surjective submersion, the pullback map $p^*: \Omega^{k}(Q)\to\Omega^{k}(G)$ is injective:  indeed, for each $x\in Q$, there exists an open neighborhood $U\subseteq Q$ and a smooth section $s: U\to G$ with $p\circ s=\mathrm{id}_U$, and then for any $\beta\in\Omega^{k}(Q)$, \begin{align*}
\beta\vert_U  =  (\mathrm{id}_U)^*\beta =  (p\circ s)^*\beta =  s^*(p^*\beta).
\end{align*} Thus, if $p^*\beta=0$, then $\beta\vert_U=0$ for all such $U$. Hence, $\beta=0$ globally.
Applying this to $\beta=d\beta_\nu$ gives $d\beta_\nu  =  0$.  This completes the proof of the proposition.
\end{proof}

\begin{theorem}
\label{thm:twistedcota}
Let $M$ be a smooth manifold. Let $\pi:T^*M\to M$ be the bundle projection, and let  $\omega_{\mathrm{can}} = - d\vartheta_{M}$ be the canonical symplectic form on $T^*M$, where $\vartheta_{M}$ is the tautological one form. Let $\sigma\in\Omega^2(M)$ be a closed $2$-form and let $\varepsilon\in\mathbb{R}$.
%Let $N = K_1 \backslash G/K_2$ be smooth (i.e., the $H=K_1\times K_2$ action on $G$ is free and proper).
%Let $\pi: T^*M\to M$ be the bundle projection and let $\omega_{\mathrm{can}}=-d\vartheta_{M}$ be the canonical symplectic form on $T^*M$.
Define \begin{align*} 
\omega_{\varepsilon}:= \omega_{\mathrm{can}}  +  \varepsilon\,\pi^*\sigma  \in \Omega^2(T^*M).
\end{align*} Then $\omega_{\varepsilon}$ is symplectic on $T^*M$ for every $\varepsilon\in\mathbb{R}$.
In particular, we may take $\sigma = \beta_\nu$ arising from Proposition \ref{prop:curvature}. Then $\omega_{\varepsilon}$ is symplectic for all $\varepsilon\in\mathbb{R}$.
\end{theorem}

\begin{proof}
We prove that $\omega_{\varepsilon}$ is closed and nondegenerate.

\medskip 

We first show that $\omega_\varepsilon$ is closed. Since $\omega_{\mathrm{can}}= - d\vartheta_{M}$, we have $d\omega_{\mathrm{can}}=0$. Since $\sigma$ is closed, $d\sigma=0$. Using functoriality of pullback and linearity of $d$, a simple computation shows \begin{align*}
d\omega_{\varepsilon} =  d\bigl(\omega_{\mathrm{can}} + \varepsilon\,\pi^*\sigma\bigr) %\\
%&=  d\omega_{\mathrm{can}}  +  \varepsilon\,d(\pi^*\sigma) \\ 
%&=  0  +  \varepsilon\,\pi^*(d\sigma)\\
 =  \varepsilon\,\pi^*(0) %\\
 =  0.
\end{align*} Hence, $\omega_{\varepsilon}$ is closed.

We then show that $\omega_\varepsilon$ is nondegenerate. Fix a point $\alpha_q \in T^*M$ and choose canonical cotangent coordinates $(q^i,p_i)$ induced by a coordinate chart on $M$ such that $\omega_{\mathrm{can}} = \sum_i dq^i \wedge dp_i$. Since $\sigma$ is a form on the base $M$, $\pi^*\sigma$ only involves $dq^i \wedge dq^j$ terms. In the ordered basis $\{\partial_{q^i},\partial_{p_i}\}$, the matrix of $\omega_\varepsilon$ (with entries $\omega_\varepsilon(e_a,e_b)$) is the block matrix $\left(\begin{smallmatrix}
\varepsilon(\sigma_{ij}) & I\\
-I & 0
\end{smallmatrix}\right)$, where $\sigma_{ij}:=\sigma(\partial_{q^i},\partial_{q^j})$. The invertible off-diagonal blocks guarantee that $\omega_\varepsilon$ is everywhere nondegenerate.

We have shown $d\omega_{\varepsilon}=0$ and that $\omega_{\varepsilon}$ is nondegenerate.
Therefore, $\omega_{\varepsilon}$ is a symplectic form on $T^*M$. Finally, if $\sigma=\beta_\nu$ arises from Proposition \ref{prop:curvature}, then $d\beta_\nu=0$ has already been proved.
\end{proof}

\subsection{The magnetic twisting on $T^*(G/T)$}

For a cotangent bundle $T^*N$, we write $\vartheta_{N}$ for its tautological one-form and use the convention $\omega_{\mathrm{can},N}:=-d\vartheta_{N}$.

Let $G$ be a Lie group, and let $T\subseteq G$ be a closed subgroup.
Fix $\lambda \in \mathfrak{g}^*$ whose stabilizer for the coadjoint action is exactly $T$ such that the coadjoint orbit $\mathcal{O}_\lambda:= G\cdot\lambda$ is diffeomorphic to $G/T$.
Define the orbit map \begin{align*}
\iota_\lambda :  G/T \longrightarrow \mathcal{O}_\lambda,\qquad \iota_\lambda(gT)=\mathrm{Ad}^*_g\lambda.
\end{align*}

\begin{definition} 
The KKS $2$-form $\omega_{\mathrm{KKS},\lambda}$ on $\mathcal{O}_\lambda$ is defined by \begin{align*}
\big(\omega_{\mathrm{KKS},\lambda}\big)_\xi\big( \mathrm{ad} _X^*\xi, \mathrm{ad}_Y^*\xi\big) := \langle \xi,[X,Y]\rangle, \qquad \xi\in\mathcal{O}_\lambda,\;
X,Y\in\mathfrak{g},
\end{align*} where $\mathrm{ad}^*_X\xi$ denotes the fundamental vector at $\xi$ for the infinitesimal coadjoint action.
\end{definition}

\begin{lemma}
\label{lem:KKSbra}
The $2$-form $\omega_{\mathrm{KKS},\lambda}$ is well-defined, $G$-invariant, closed, and nondegenerate. Consequently, $\sigma_\lambda:=\iota_\lambda^*\omega_{\mathrm{KKS},\lambda}\in\Omega^2(G/T)$ is a symplectic form on $G/T$.
\end{lemma}

\begin{proof}
These standard KKS properties are discussed in \cite{Cannas2001}. The KKS form is intrinsically well-defined and $G$-invariant by construction: if $\mathrm{ad}_X^*\xi=0$, then $\langle\xi,[X,Y]\rangle=0$ for all $Y$, so the defining pairing depends only on the tangent vectors $\mathrm{ad}_X^*\xi,\mathrm{ad}_Y^*\xi$.
Its closedness $d\omega_{\mathrm{KKS},\lambda} = 0$ follows directly from the Jacobi identity of the Lie algebra $\mathfrak{g}$ applied to Cartan's formula. Nondegeneracy is immediate since $\langle \xi, [X,Y] \rangle = 0$ for all $Y$ implies $\mathrm{ad}_X^* \xi = 0$, meaning the fundamental vector field vanishes.
\end{proof}

For the application in the main text, return to compact, connected, semisimple $G$ and $\lambda=B^\flat(\gamma)$ with regular $\gamma$, so that $T=Z_G(\gamma)$. The twisting form is the negative KKS pullback, \begin{align*}
\sigma_{\gamma} := -\iota_\lambda^*\omega_{\mathrm{KKS},\lambda} = -\sigma_\lambda.
\end{align*} Let $\pi: T^*(G/T)\to G/T$ be the bundle projection, and let $\omega_{\mathrm{can}}=-d\vartheta_{G/T}$ be the canonical symplectic form.
Fix $\varepsilon\in\mathbb{R}$ and define the magnetic (twisted cotangent) form
\begin{equation}
\omega_\varepsilon := \omega_{\mathrm{can}} + \varepsilon\,\pi^*\sigma_{\gamma} \in \Omega^2\bigl(T^*(G/T)\bigr). \label{eq:omegaeps}
\end{equation}

\begin{proposition}
\label{prop:twistomeg}
For every $\varepsilon\in\mathbb{R}$, the $2$-form $\omega_\varepsilon$ is symplectic on $T^*(G/T)$.
Moreover, $\omega_0=\omega_{\mathrm{can}}$ is the non-twisted cotangent symplectic form.
\end{proposition}

\begin{proof}
This is an immediate consequence of Theorem \ref{thm:twistedcota} applied to $M=G/T$ equipped with the closed $2$-form $\sigma=\sigma_{\gamma}\in\Omega^2(G/T)$.
\end{proof}

   \bibliographystyle{unsrt}
\bibliography{bibliography.bib}

@article{jiang2026poisson,
  title   = {Poisson Centralisers and Polynomial Superintegrability for Magnetic Geodesic Flows on Reductive Homogeneous Spaces},
  author  = {Jiang, K. and Ma, G. and Marquette, I. and Zhang, J. and Zhang, Y.-Z.},
  journal = {arXiv preprint arXiv:2601.01369},
  year    = {2026}
}

@article {MR3119484,
    AUTHOR = {Miller Jr, W. and Post, S. and Winternitz, P.},
     TITLE = {Classical and quantum superintegrability with applications},
   JOURNAL = {J. Phys. A: Math. Theor.},
  FJOURNAL = {Journal of Physics. A. Mathematical and Theoretical},
    VOLUME = {46},
      YEAR = {2013},
    NUMBER = {42},
     PAGES = {423001, 97},
      ISSN = {1751-8113},
   MRCLASS = {70H06 (81R12)},
  MRNUMBER = {3119484},
       DOI = {10.1088/1751-8113/46/42/423001},
       URL = {https://doi-org.wwwproxy1.library.unsw.edu.au/10.1088/1751-8113/46/42/423001},
}

@article{nehorovsev1968action,
  title   = {Action-angle variables and their generalizations},
  author  = {Nehoro{\v{s}}ev, N. N.},
  journal = {Trudy Moskov. Mat. Obshch.},
  volume  = {26},
  year    = {1972},
  pages   = {181--198}
}

@article {MR187763,
    AUTHOR = {Fri\v{s}, J. and Mandrosov, V. and Smorodinsky, Ya. A. and Uhl\'{\i}\v{r},
              M. and Winternitz, P.},
     TITLE = {On higher symmetries in quantum mechanics},
   JOURNAL = {Phys. Lett.},
  FJOURNAL = {Physics Letters},
    VOLUME = {16},
      YEAR = {1965},
     PAGES = {354--356},
      ISSN = {0031-9163},
   MRCLASS = {81.99},
  MRNUMBER = {187763},
MRREVIEWER = {J. M. Jauch},
}

@incollection {MR4285685,
    AUTHOR = {Reshetikhin, N.},
     TITLE = {Spin {C}alogero-{M}oser models on symmetric spaces},
 BOOKTITLE = {Integrability, quantization, and geometry. {I}. {I}ntegrable
              systems},
    SERIES = {Proc. Sympos. Pure Math.},
    VOLUME = {103.1},
     PAGES = {377--402},
 PUBLISHER = {Amer. Math. Soc., Providence, RI},
      YEAR = {[2021] \copyright 2021},
      ISBN = {978-1-4704-5591-0},
   MRCLASS = {53D05 (37J37)},
  MRNUMBER = {4285685},
       DOI = {10.1090/pspum/103.1/01840},
       URL = {https://doi.org/10.1090/pspum/103.1/01840},
}

@article{Hochgerner2008SingularCotangent,
  author  = {Hochgerner, S.},
  title   = {Singular cotangent bundle reduction \& spin Calogero--Moser systems},
  journal = {Differential Geometry and its Applications},
  volume  = {26},
  number  = {2},
  year    = {2008},
  pages   = {169--192},
  doi     = {10.1016/j.difgeo.2007.11.008}
}

@book{Lee2013IntroductionSmoothManifolds,
  author    = {Lee, John M.},
  title     = {Introduction to Smooth Manifolds},
  series    = {Graduate Texts in Mathematics},
  volume    = {218},
  edition   = {2},
  publisher = {Springer},
  address   = {New York},
  year      = {2013},
  doi       = {10.1007/978-1-4419-9982-5},
  isbn      = {978-1-4419-9981-8}
}

@book{marsden2007hamiltonian,
  title = {Hamiltonian Reduction by Stages},
  author = {Marsden, Jerrold E. and Misiolek, Gerard and Ortega, Juan-Pablo and Perlmutter, Matthew and Ratiu, Tudor S.},
  series = {Lecture Notes in Mathematics},
  volume = {1913},
  year = {2007},
  publisher = {Springer},
  doi = {10.1007/978-3-540-72470-4}
}

@article {MR3493688,
    AUTHOR = {Reshetikhin, N.},
     TITLE = {Degenerately integrable systems},
   JOURNAL = {Zap. Nauchn. Sem. S.-Peterburg. Otdel. Mat. Inst. Steklov.
              (POMI)},
  FJOURNAL = {Rossi\u{\i}skaya Akademiya Nauk. Sankt-Peterburgskoe Otdelenie.
              Matematicheski\u{\i} Institut im. V. A. Steklova. Zapiski Nauchnykh
              Seminarov (POMI)},
    VOLUME = {433},
      YEAR = {2015},
    NUMBER = {Voprosy Kvantovo\u{\i} Teorii Polya i Statistichesko\u{\i} Fiziki. 23},
     PAGES = {224--245},
      NOTE = {English transl.: \emph{J. Math. Sci. (N.Y.)} \textbf{219} (2016), no. 6, 890--909},
      ISSN = {0373-2703},
   MRCLASS = {70H06 (17B63 17B80 37K10)},
  MRNUMBER = {3493688},
MRREVIEWER = {Luca Degiovanni},
       DOI = {10.1007/s10958-016-2738-9},
       URL = {https://doi.org/10.1007/s10958-016-2738-9},
}

@article {MR4700605,
    AUTHOR = {Reshetikhin, N.},
     TITLE = {Spin {C}alogero-{M}oser periodic chains and two dimensional
              {Y}ang-{M}ills theory with corners},
   JOURNAL = {Pure Appl. Math. Q.},
  FJOURNAL = {Pure and Applied Mathematics Quarterly},
    VOLUME = {19},
      YEAR = {2023},
    NUMBER = {5},
     PAGES = {2537--2572},
      ISSN = {1558-8599,1558-8602},
   MRCLASS = {81R12 (37J35 81T13)},
  MRNUMBER = {4700605},
}

@article {MR4071113,
    AUTHOR = {Fournier, F. and {\v S}nobl, L. and Winternitz, P.},
     TITLE = {Cylindrical type integrable classical systems in a magnetic
              field},
   JOURNAL = {J. Phys. A},
  FJOURNAL = {Journal of Physics. A. Mathematical and Theoretical},
    VOLUME = {53},
      YEAR = {2020},
    NUMBER = {8},
     PAGES = {085203, 31},
      ISSN = {1751-8113,1751-8121},
   MRCLASS = {37J35 (70H06)},
  MRNUMBER = {4071113},
MRREVIEWER = {Xiaoxue\ Xu},
       DOI = {10.1088/1751-8121/ab64a6},
       URL = {https://doi.org/10.1088/1751-8121/ab64a6},
}

@book {MR3942135,
    AUTHOR = {Kalnins, E. G. and Kress, J. M. and Miller, Jr.,
              W.},
     TITLE = {Separation of variables and superintegrability},
    SERIES = {IOP Expanding Physics},
      NOTE = {The symmetry of solvable systems},
 PUBLISHER = {IOP Publishing, Bristol},
      YEAR = {2018},
     PAGES = {xv+approximately 300},
      ISBN = {978-0-7503-1314-8; 978-0-7503-1315-5},
   MRCLASS = {35-02 (35A22 35A25 37Jxx)},
  MRNUMBER = {3942135},
MRREVIEWER = {Giovanni\ Rastelli},
       DOI = {10.1088/978-0-7503-1314-8},
       URL = {https://doi.org/10.1088/978-0-7503-1314-8},
}

@article {MR0402819,
    AUTHOR = {Marsden, J. and Weinstein, A.},
     TITLE = {Reduction of symplectic manifolds with symmetry},
   JOURNAL = {Rep. Mathematical Phys.},
  FJOURNAL = {Reports on Mathematical Physics},
    VOLUME = {5},
      YEAR = {1974},
    NUMBER = {1},
     PAGES = {121--130},
      ISSN = {0034-4877},
   MRCLASS = {58F05 (70.58 76.58)},
  MRNUMBER = {402819},
MRREVIEWER = {K.\ Lamotke},
       DOI = {10.1016/0034-4877(74)90021-4},
       URL = {https://doi.org/10.1016/0034-4877(74)90021-4},
}

@book{Cannas2001,
  author    = {Cannas da Silva, Ana},
  title     = {Lectures on Symplectic Geometry},
  series    = {Lecture Notes in Mathematics},
  volume    = {1764},
  publisher = {Springer},
  address   = {Berlin},
  year      = {2001},
  doi       = {10.1007/978-3-540-45330-7}
}

@article{Bolsinov2008,
  author  = {Bolsinov, Alexey V. and Jovanovi{\'c}, Bo{\v z}idar},
  title   = {Magnetic flows on homogeneous spaces},
  journal = {Commentarii Mathematici Helvetici},
  volume  = {83},
  number  = {3},
  year    = {2008},
  pages   = {679--700},
  doi     = {10.4171/CMH/139}
}

@article{MishchenkoFomenkoGeneralized1978,
  author = {Mishchenko, A. S. and Fomenko, A. T.},
  title = {Generalized {Liouville} method of integration of {Hamiltonian} systems},
  journal = {Functional Analysis and Its Applications},
  volume = {12},
  number = {2},
  pages = {113--121},
  year = {1978},
  doi = {10.1007/BF01076254}
}

@article{Kummer1981Reduction,
  author = {Kummer, M.},
  title = {On the construction of the reduced phase space of a {Hamiltonian} system with symmetry},
  journal = {Indiana University Mathematics Journal},
  volume = {30},
  number = {2},
  pages = {281--291},
  year = {1981},
  doi = {10.1512/iumj.1981.30.30022}
}

@article{MarsdenPerlmutter2000Orbit,
  author = {Marsden, Jerrold E. and Perlmutter, Matthew},
  title = {The orbit bundle picture of cotangent bundle reduction},
  journal = {Comptes Rendus Math{\'e}matiques de l'Acad{\'e}mie des Sciences, La Soci{\'e}t{\'e} Royale du Canada},
  volume = {22},
  number = {2},
  pages = {33--54},
  year = {2000}
}

@article{Feher2023Master,
  author = {Feh{\'e}r, L{\'a}szl{\'o}},
  title = {Poisson reductions of master integrable systems on doubles of compact {Lie} groups},
  journal = {Annales Henri Poincar{\'e}},
  volume = {24},
  pages = {1823--1876},
  year = {2023},
  doi = {10.1007/s00023-022-01260-3},
  eprint = {2208.03728},
  archivePrefix = {arXiv}
}

@misc{Feher2024Degenerate,
  author = {Feh{\'e}r, L{\'a}szl{\'o}},
  title = {Notes on the degenerate integrability of reduced systems obtained from the master systems of free motion on cotangent bundles of compact {Lie} groups},
  year = {2024},
  eprint = {2309.16245},
  archivePrefix = {arXiv},
  primaryClass = {math-ph},
  note = {arXiv:2309.16245, version 3}
}
\end{document}